\documentclass{article}
\usepackage{amsmath,amsfonts,graphicx}
\usepackage{amssymb}
\usepackage{amsthm}
\usepackage{url}
\usepackage[numbers,square,sort&compress]{natbib}
\usepackage{algorithm}
\usepackage{algorithmic}
\usepackage{psfrag}
\usepackage[utf8]{inputenc}

\usepackage{bbold}

\newcommand{\kpk}{_{k+1|k}}
\newcommand{\kmkm}{_{k-1|k-1}}

\newcommand{\kkm}{_{k|k-1}}

\renewcommand{\k}{_k}
\newcommand{\kk}{_{k|k}}
\newcommand{\kp}{_{k+1}}
\newcommand{\otk}{_{1:k}}
\newcommand{\otkm}{_{1:k-1}}

\newcommand{\sui}{^{(i)}}
\newcommand{\suj}{^{(j)}}

\newcommand{\N}{\ensuremath{\mathcal{N}}} 
\newcommand{\gvn}{\,|\,} 
\newcommand{\xh}{\hat x}
\newcommand{\yh}{\hat y}
\newcommand{\set}[1]{\{ #1 \}}
\newcommand{\one}{\mathbb{1}}

\newcommand{\bpm}{\begin{pmatrix}}
\newcommand{\epm}{\end{pmatrix}}
\newcommand{\bbm}{\begin{bmatrix}}
\newcommand{\ebm}{\end{bmatrix}}
\newcommand{\inv}{^{-1}}
\newcommand{\sr}{^\frac{1}{2}}
\newcommand{\srt}{^\frac{\mathrm{T}}{2}}
\newcommand{\T}{^{T}}

\DeclareMathOperator{\diag}{diag}

\DeclareMathOperator{\cov}{cov}
\DeclareMathOperator{\E}{E}
\newcommand{\ordo}{\mathcal{O}}

\usepackage{xspace}

\title{The Ensemble Kalman Filter: A~Signal~Processing~Perspective}
\author{
Michael Roth, Gustaf Hendeby, Carsten Fritsche,\\and Fredrik Gustafsson\\
Dept. of Electrical Engineering,\\Linköping University,\\Linköping, Sweden\\ 
michael.roth@liu.se}

\begin{document}
\maketitle

\begin{abstract}
The ensemble Kalman filter (EnKF) is a Monte Carlo based implementation of the Kalman filter (KF) for extremely high-dimensional, possibly nonlinear and non-Gaussian state estimation problems. Its ability to handle state dimensions in the order of millions has made the EnKF a popular algorithm in different geoscientific disciplines. Despite a similarly vital need for scalable algorithms in signal processing, e.g., to make sense of the ever increasing amount of sensor data, the EnKF is hardly discussed in our field. 



This self-contained review paper is aimed at signal processing researchers and provides all the knowledge to get started with the EnKF. The algorithm is derived in a KF framework, without the often encountered geoscientific terminology. Algorithmic challenges and required extensions of the EnKF are provided, as well as relations to sigma-point KF and particle filters. The relevant EnKF literature is summarized in an extensive survey and unique simulation examples, including popular benchmark problems, complement the theory with practical insights. The signal processing perspective highlights new directions of research and facilitates the exchange of potentially beneficial ideas, both for the EnKF and high-dimensional nonlinear and non-Gaussian filtering in general. 
%
%
%
\end{abstract}


%



\section{Introduction}
\label{sec:intro}
Numerical weather prediction~\cite{kalnay_atmospheric_2002} is an extremely high-dimensional geoscientific state estimation problem. 
The state~$x$ comprises physical quantities (temperature, wind speed, air pressure, etc.) at many spatially distributed grid points, which often yields a state dimension $n$ in the order of millions. 
%
%
Consequently, the Kalman filter (KF)~\cite{kalman_new_1960, anderson_optimal_1979} or its nonlinear extensions~\cite{julier_new_1995,roth_nonlinear_2016} that require the storage and processing of $n\times n$ covariance matrices cannot be applied directly. It is well-known that the application of particle filters~\cite{gordon_novel_1993,gustafsson_particle_2010} is not feasible either. 
In contrast, the ensemble Kalman filter~\cite{evensen_sequential_1994,burgers_analysis_1998} (EnKF) was specifically developed as algorithm for high-dimensional $n$. 
The EnKF
\begin{itemize}
\item is a random-sampling implementation of the KF;
\item reduces the computational complexity of the KF by propagating an ensemble of $N<n$ state realizations;
\item can be applied to nonlinear state-space models without the need to compute Jacobian matrices;
\item can be applied to continuous-time as well as discrete-time state transition functions;
\item can be applied to non-Gaussian noise densities;
\item is simple to implement;
\item does not converge to the Bayesian filtering solution for $N\rightarrow\infty$ in general;
\item often requires extra measures to work in practice. 
\end{itemize}

Also in the field of stochastic signal processing (SP) and Bayesian state estimation, high-dimensional problems become more and more relevant. Examples include SLAM~\cite{durrant-whyte_simultaneous_2006} where $x$ contains an increasing number of landmark positions, or extended target tracking~\cite{baum_extended_2014,wahlstrom_extended_2015} where $x$ can contain many parameters to describe the shape of the target. Furthermore, scalable SP algorithms are required to make sense of the ever increasing amount of data from sensors in everyday devices.
%
%
EnKF approaches hardly appear in the relevant SP journals, though. In contrast, vivid theoretical development is documented in geoscientific journals under the umbrella term data assimilation~(DA)~\cite{kalnay_atmospheric_2002}. Hence, a relevant SP problem is being addressed with only little participation from the SP community. Conversely, much of the DA literature makes little reference to relevant SP contributions. It is our intention to bridge this interesting gap.

There is further overlap that motivates for a closer investigation of the EnKF. First, the basic EnKF~\cite{burgers_analysis_1998} can be applied to nonlinear and non-Gaussian state-space models because it is entirely sampling based. In fact, the state evolution in geoscientific applications is typically governed by large nonlinear black box prediction models derived from partial differential equations. Furthermore, satellite measurements in weather applications are nonlinearly related to the states~\cite{kalnay_atmospheric_2002}. Hence, the EnKF has long been investigated as nonlinear filter. Second, the EnKF literature contains so called localization methods~\cite{houtekamer_data_1998, houtekamer_sequential_2001} to systematically approach high-dimensional problems by only acting on a part of the state vector in each measurement update. These ideas can be directly transferred to sigma point filters~\cite{roth_nonlinear_2016}. Third, the EnKF offers several interesting opportunities to apply SP techniques, e.g., via the application of bootstrap or regularization methods in the EnKF gain computation. 

The contributions of this paper aim at making the EnKF more accessible to SP researchers. 
We provide a concise derivation of the EnKF based on the KF. A literature review highlights important EnKF papers with their respective contributions, and facilitates an easier access to the extensive and rapidly developing DA literature on the EnKF. Moreover, we put the EnKF in context with popular SP algorithms such as sigma point filters~\cite{julier_new_1995,roth_nonlinear_2016} and the particle filter~\cite{gordon_novel_1993,gustafsson_particle_2010}. Our presentation forms a solid basis for further developments and the transfer of beneficial ideas and techniques between the fields of SP and DA.

The structure of the paper is as follows. After an extensive literature review in Sec.~\ref{sec:literature}, the EnKF is developed from the KF in Sec.~\ref{sec:enkf}. Algorithmic properties and challenges of the EnKF and the available approaches to face them are discussed in Sec.~\ref{sec:enkfProp} and~\ref{sec:enkfExt}, respectively. Relations to other filtering algorithms are discussed in Sec.~\ref{sec:enkfRelations}. The theoretical considerations are followed by numerical simulations in Sec.~\ref{sec:examples} and some concluding remarks in Sec.~\ref{sec:conclusion}.


\section{Filtering and EnKF literature}
\label{sec:literature}

The following literature review provides important landmarks for the EnKF novice. 

State-space models and the filtering problem have been investigated since the 1960s. Early results include the Kalman filter (KF)~\cite{kalman_new_1960} as algorithm for linear systems, and the Bayesian filtering equations~\cite{jazwinski_stochastic_1970} as theoretical solution for nonlinear and non-Gaussian systems. Because the latter approach cannot be implemented in general, approximate filtering algorithms are required. With a leap in computing capacity, the 1990s saw major developments. The sampling-based sigma point Kalman filters~\cite{julier_new_1995,roth_nonlinear_2016} started to appear. Furthermore, particle filters~\cite{gordon_novel_1993, gustafsson_particle_2010} were developed to approximately implement the Bayesian filtering equations through sequential importance sampling. 

The first EnKF~\cite{evensen_sequential_1994} was proposed in a geoscientific journal in 1994 and introduced the idea of propagating ensembles to mimic the KF. A systematic error that resulted in an underestimated uncertainty was later corrected by processing ``perturbed measurements''. This randomization is well motivated in~\cite{burgers_analysis_1998} but also used in~\cite{houtekamer_data_1998}. 

Interestingly,~\cite{evensen_sequential_1994} remains the most cited EnKF paper\footnote{With over 3000 citations between 1994 and 2016.}, followed by the overview article~\cite{evensen_ensemble_2003} and the monograph~\cite{evensen_data_2009} by the same author. Other insightful overviews from a geoscientific perspective are~\cite{hamill_ensemble-based_2006,houtekamer_ensemble_2005}. Many practical aspects of operational EnKF for weather prediction and re-analysis are described in~\cite{houtekamer_ensemble_2005,whitaker_ensemble_2008,compo_twentieth_2011}. Whereas the aforementioned papers were mostly published in geoscientific outlets, a special issue of the IEEE Control Systems Magazine appeared with review articles~\cite{lakshmivarahan_ensemble_2009,anderson_ensemble_2009,evensen_ensemble_2009} and an EnKF case study~\cite{mandel_data_2009}. 
%
%
Still, the above material was written by EnKF researchers with a geoscientific focus and in the application specific terminology. 
Furthermore, references to the recent SP literature and other nonlinear KF variants~\cite{roth_nonlinear_2016} are scarce. 

Some attention has been devoted to the EnKF also beyond the geosciences. Convergence properties for $N\rightarrow\infty$ have been established using different theoretical analyses of the EnKF~\cite{furrer_estimation_2007,butala_asymptotic_2008,mandel_convergence_2011}. Statistical perspectives are provided in the thesis~\cite{frei_ensemble_2013} and the review~\cite{katzfuss_understanding_2016}. A recommended SP view that connects the EnKF with Bayesian filtering and particle methods, including convergence results for nonlinear systems, is~\cite{crisan_large_2011}. 
Examples of the EnKF as tool for tomographic imaging and target tracking are described in~\cite{butala_tomographic_2009} and \cite{dunik_random-point-based_2015}, respectively.
Brief introductory papers that connect the EnKF with more established SP algorithms include~\cite{gillijns_what_2006} and~\cite{roth_ensemble_2015}. The latter also served as basis for this article.

The majority of EnKF advances is still documented in geoscientific publications. Notable contributions include deterministic EnKF that avoid the randomization of~\cite{burgers_analysis_1998} and propagate an ensemble of deviations from the ensemble mean~\cite{anderson_ensemble_2001,bishop_adaptive_2001,whitaker_ensemble_2002,evensen_ensemble_2003}. Their common basis as square root EnKF and the relation to square root KF~\cite{anderson_optimal_1979} is discussed in~\cite{tippett_ensemble_2003}. The computational advantages in high-dimensional EnKF with small ensembles ($N\ll n$) come at the price of adverse effects, including the risk of filter divergence. 
The often encountered underestimation of uncertainty can be counteracted with ensemble inflation~\cite{anderson_monte_1999}. 
A scheme with two EnKF in parallel that provide each other with gain matrices to reduce unwanted ``inbreeding'' has been suggested in~\cite{houtekamer_data_1998}. The benefit of such a double EnKF is, however, debated~\cite{van_leeuwen_comment_1999,whitaker_ensemble_2002}.
The low-rank approximation of covariance matrices can yield spurious correlations between supposedly uncorrelated state components and measurements. Localization techniques such as local measurement updates~\cite{houtekamer_data_1998,evensen_ensemble_2003,ott_local_2004} or covariance tapering~\cite{houtekamer_sequential_2001,hamill_distance-dependent_2001} let the measurement only affect a part of the state vector. In other words, localization effectively reduces the dimension of each measurement update. Inflation and localization are essential components of operational EnKF~\cite{houtekamer_ensemble_2005}. A list of further advances in the geoscientific literature is provided in the appendix of~\cite{evensen_data_2009}.

An interesting development for SP researchers is the reconsideration of particle filters (PF) for high-dimensional geoscientific problems, with seemingly little reference to SP literature. An early example is~\cite{van_leeuwen_variance-minimizing_2003}. The well-known challenges, mostly related to the problem of importance sampling in high dimensions, are reviewed in~\cite{snyder_obstacles_2008, van_leeuwen_particle_2009}. Several recent approaches~\cite{van_leeuwen_nonlinear_2010,frei_bridging_2013,poterjoy_localized_2015} were successfully tested on a popular EnKF benchmark problem~\cite{lorenz_predictability_2006} that is also investigated in the simulation examples of this paper. Combinations of the EnKF with the deterministic sampling of sigma point filters~\cite{roth_nonlinear_2016} are given in~\cite{pham_stochastic_2001} and \cite{luo_ensemble_2009}. However, the benefit of the unscented transformation~\cite{julier_unscented_2004,roth_nonlinear_2016} in~\cite{luo_ensemble_2009} is debated in~\cite{sakov_comment_2009}. Ideas to combine the EnKF with Gaussian mixture approaches are given in~\cite{stordal_bridging_2011,hoteit_particle_2011,frei_mixture_2013}.

\section{A Signal Processing Introduction to the Ensemble Kalman Filter}
\label{sec:enkf}

The underlying framework of our filter presentation are discrete-time state-space models~\cite{anderson_optimal_1979,jazwinski_stochastic_1970}. 
The Kalman filter and many EnKF variants are built upon the linear model
\begin{subequations}\label{eq:ssm}
\begin{align}
x\kp &= F x\k + G v\k, \label{eq:stateDiff}\\
y\k &= H x\k + e\k, \label{eq:meas}
\end{align}
\end{subequations}
with the $n$-dimensional state $x$ and the $m$-dimensional measurement $y$. The initial state $x_0$, the process noise $v\k$, and the measurement noise $e\k$ are described by $\E(x_0)=\xh_0$, $\E(v\k)=0$, $\E(e\k)=0$, $\cov(x_0)=P_0$, $\cov(v\k)=Q$, and $\cov(e\k)=R$. In the Gaussian case, these moments completely characterize the distributions of $x_0$, $v\k$, and $e\k$.

Nonlinear relations in the state evolution and measurement equations can be described by a more general model
\begin{subequations}\label{eq:ssmNonlin}
\begin{align}
x\kp &= f(x\k,v\k), \label{eq:stateDiffNonlin}\\
y\k &= h(x\k, e\k). \label{eq:measNonlin}
\end{align}
\end{subequations}
More general noise and initial state distributions can, for example, be characterized by probability density functions $p(x_0)$, $p(v\k)$, and $p(e\k)$. 

Both \eqref{eq:ssm} and \eqref{eq:ssmNonlin} can be time-varying but the time indices for functions and matrices are omitted for convenience. 

\subsection{A brief Kalman filter review}\label{sec:KF}

The KF is an optimal linear filter~\cite{anderson_optimal_1979} for~\eqref{eq:ssm} that propagates state estimates $\xh\kk$ and covariance matrices~$P\kk$.

The KF time update or prediction is given by
\begin{subequations}\label{eq:kfTime}
\begin{align}
\xh\kpk &= F \xh\kk,  \label{eq:kfTimeState}\\
P\kpk &= F P\kk F\T + G Q G\T. \label{eq:kfTimeCov}
\end{align}
\end{subequations}
The above parameters can be used to predict the output of~\eqref{eq:ssm} and its uncertainty via
\begin{subequations}\label{eq:kfOutput}
\begin{align}
\yh\kkm &= H \xh\kkm, \label{eq:kfOutputHat}\\
S\k &= H P\kkm H\T + R. \label{eq:kfResCov}
\end{align}
\end{subequations}

The measurement update adjusts the prediction results according to
\begin{subequations}\label{eq:kfMeas}
\begin{align}
\xh\kk &= \xh\kkm + K\k (y\k - \yh\kkm) \label{eq:kfMeasState1}\\
 &=(I-K\k H) \xh\kkm + K\k y\k, \label{eq:kfMeasState2}\\
P\kk &= (I-K\k H) P\kkm (I-K\k H)\T + K\k R K\k\T, \label{eq:kfMeasCov}
\end{align}
\end{subequations}
with a gain matrix $K\k$. Here, \eqref{eq:kfMeasState2} resembles a deterministic observer and only requires all eigenvalues of $(I-K\k H)$ inside the unit circle to obtain a stable filter. 
The optimal $K\k$ in the minimum variance sense is given by
\begin{align}
\label{eq:kfGain}
K\k &= P\kkm H\T S\k\inv = M\k S\k\inv,
\end{align}
where $M\k$ is the cross-covariance between the state and output predictions. Alternatives to the covariance update~\eqref{eq:kfMeasCov} exist, but the shown Joseph form~\cite{anderson_optimal_1979} will simplify the derivation of the EnKF. Furthermore, it is valid for all gain matrices $K\k$ beyond \eqref{eq:kfGain} and numerically well-behaved. It should be noted that it is numerically advisable to obtain $K\k$ by solving $K\k S\k = M\k$ rather than explicitly computing $S\k\inv$~\cite{trefethen_numerical_1997}.





\subsection{The ensemble idea}


The central idea of the EnKF is to propagate an ensemble of $N<n$ (often $N\ll n$) state realizations $\smash{\set{x\sui\k}_{i=1}^N}$ instead of the $n$-dimensional estimate $\xh\kk$ and the $n\times n$ covariance $P\kk$ of the KF. The ensemble is processed such that 
%
%
\begin{subequations}\label{eq:sampleStat}
\begin{align}
\bar x\kk &= \tfrac{1}{N}\sum\nolimits_{i=1}^N x\sui\k \approx  \xh\kk,\\
\bar P\kk &= \tfrac{1}{N-1}\sum\nolimits_{i=1}^N \bigl(x\sui\k-\bar x\kk\bigr)\bigl(x\sui\k-\bar x\kk\bigr)\T \label{eq:sampleCov}
\approx P\kk.
\end{align}
\end{subequations}
Reduced computational complexity is achieved because the explicit computation of $\bar P\kk$ is avoided in the EnKF recursion.
%
%
Of course, this reduction comes at the price of a low rank approximation in~\eqref{eq:sampleCov} that entails some negative effects and requires extra measures.

For our development it is convenient to treat the ensemble as an $n\times N$ matrix $X\kk$ with columns $x\sui\k$. This allows for the compact notation of the ensemble mean and covariance
\begin{subequations}\label{eq:encode}
\begin{align}
\bar x\kk &= \tfrac{1}{N}X\kk\one, \label{eq:sampleMean}\\
\bar P\kk &= \tfrac{1}{N-1}\widetilde X\kk \widetilde X\kk\T,  \label{eq:sampleCov2}
\end{align}
\end{subequations}
where $\one=[1, \hdots, 1]\T$ is an $N$-dimensional vector and 
%
\begin{equation}
\widetilde X\kk = X\kk - \bar x\kk\one\T = X\kk (I_N - \tfrac{1}{N} \one\one\T) \label{eq:ensDev}
\end{equation}
%
is an ensemble of deviations from $\bar x\kk$, sometimes called ensemble anomalies~\cite{evensen_data_2009}. The matrix multiplication in~\eqref{eq:ensDev} provides a compact way to write the anomalies, but is not the most efficient way to compute them.





\subsection{The EnKF time update}


The EnKF time update is referred to as forecast in the geoscientific literature. In analogy to~\eqref{eq:kfTime}, a prediction ensemble $X\kpk$ is computed that carries the information in $\xh\kpk$ and $P\kpk$. An ensemble of $N$ independent process noise realizations $\smash{\set{v\sui\k}_{i=1}^N}$ with zero mean and covariance $Q$, stored as matrix $V\k$, is used in
\begin{equation}
X\kpk = F X\kk + G V\k.
\label{eq:enkfPred}
\end{equation}
An extension to nonlinear state transitions \eqref{eq:stateDiffNonlin} is given by
\begin{equation}
X\kpk = f(X\kk, V\k),\label{eq:enkfPredNonlin}
\end{equation}
where we generalized $f$ to act on the columns of its input matrices. Apparently, the EnKF time update amounts to a one-step-ahead simulation of $X\kk$. Consequently, also continuous-time dynamics can be considered by, for example, numerically solving partial differential equations to obtain $X\kpk$. Also non-Gaussian initial state and process noise distributions with arbitrary densities $p(x_0)$ and $p(v\k)$ can be employed as long as they allow sampling. Perhaps because of this flexibility, the time update is often omitted in the EnKF literature~\cite{burgers_analysis_1998, houtekamer_data_1998}.


\subsection{The EnKF measurement update}


The EnKF measurement update is referred to as analysis in the geoscientific literature. A prediction or forecast ensemble $X\kkm$ is processed to obtain the filtering ensemble $X\kk$ that encodes the KF mean and covariance. We assume that a gain $\bar K\k=K\k$ is given and postpone its computation to the next section. 

With $\bar K\k$ available, the KF update~\eqref{eq:kfMeasState2} can be applied to each ensemble member as follows~\cite{evensen_sequential_1994}
\begin{equation}
X\kk = (I-\bar K\k H) X\kkm + \bar K\k y\k \one\T.
\end{equation}
The resulting ensemble average \eqref{eq:sampleMean} approximates the correct $\xh\kk$. However, with $y\k \one\T$ known, the sample covariance~\eqref{eq:sampleCov2} gives only the first term of \eqref{eq:kfMeasCov} and therefore fails to resemble $P\kk$.
A solution~\cite{burgers_analysis_1998} is to account for the missing term $\bar K\k R \bar K\k\T$ by adding artificial zero-mean measurement noise realizations $\smash{\set{e\sui\k}_{i=1}^N}$ with covariance $R$, stored as matrix $E\k$, 
according to
\begin{equation}
X\kk = (I-\bar K\k H) X\kkm + \bar K\k y\k \one\T - \bar K\k E\k. \label{eq:enkfMeas1}
\end{equation}
Then, $X\kk$ correctly resembles $\xh\kk$ and $P\kk$. 
The model \eqref{eq:ssm} is implicit in \eqref{eq:enkfMeas1} because the matrix $H$ appears. If we, in analogy to \eqref{eq:kfOutput}, define a predicted output ensemble
\begin{equation}
Y\kkm = H X\kkm + E\k
\label{eq:enkfPredMeas}
\end{equation}
that encodes $\yh\kkm$ and $S\k$, we can reformulate \eqref{eq:enkfMeas1} to an update that resembles \eqref{eq:kfMeasState1}:
\begin{equation}
X\kk = X\kkm + \bar K\k (y\k\one\T - Y\kkm). \label{eq:enkfMeas2}
\end{equation}
In contrast to~\eqref{eq:enkfMeas1}, the update~\eqref{eq:enkfMeas2} is entirely sampling based. As a consequence, we can extend the algorithm to nonlinear measurement models~\eqref{eq:measNonlin} by replacing \eqref{eq:enkfPredMeas} with
\begin{equation}
Y\kkm = h(X\kkm, E\k), \label{eq:enkfPredMeasNonlin}
\end{equation}
where we generalized $h$ to accept matrix inputs similar to~\eqref{eq:enkfPredNonlin}.

In the EnKF literature, the prevailing view of inserting artificial noise is that perturbed measurements $y\k\one\T-E\k$ are processed. This might appear unusual from an SP perspective since it suggests that information is distorted before processing. The introduction of output ensembles $Y\kkm$, in contrast, yields a direct connection to~\eqref{eq:kfOutput} and highlights the similarities between \eqref{eq:enkfMeas2} and \eqref{eq:kfMeasState1}.

An interesting point~\cite{frei_mixture_2013} is that the measurement enters linearly in~\eqref{eq:enkfMeas1} and~\eqref{eq:enkfMeas2} and merely shifts the ensemble locations. This highlights the EnKF roots in the linear KF in which $P\kk$ also remains unchanged by $y\k$.

\subsection{The EnKF gain}
\label{sec:Kcomp}


The optimal gain~\eqref{eq:kfGain} in the KF is computed from the covariance matrices of the predicted state and output. In the EnKF, the required $M\k$ and $S\k$ are not available but must be approximated from the prediction ensembles~\eqref{eq:enkfPred} or~\eqref{eq:enkfPredNonlin}, and~\eqref{eq:enkfPredMeas} or~\eqref{eq:enkfPredMeasNonlin}. 

A straightforward way to compute the EnKF gain $\bar K\k$ is to first compute the deviations or anomalies
\begin{subequations}\label{eq:msSamp}
\begin{align}
\widetilde X\kkm &= X\kkm (I_N - \tfrac{1}{N} \one\one\T),\label{eq:tildeX}\\
\widetilde Y\kkm &= Y\kkm (I_N - \tfrac{1}{N} \one\one\T),\label{eq:tildeY}
\end{align}
%
and second the sample covariance matrices
\begin{align}
\bar M\k &= \tfrac{1}{N-1}\widetilde X\kkm \widetilde Y\kkm\T,\\
\bar S\k &= \tfrac{1}{N-1}\widetilde Y\kkm \widetilde Y\kkm\T.
\end{align}
\end{subequations}
The computations \eqref{eq:msSamp} are entirely sampling-based, which is useful for the nonlinear case but introduces extra sampling errors. An obvious improvement for additive measurement noise $e\k$ with covariance $R$ is given in Sec.~\ref{sec:enkfExtSquareRoot}, together with the square root EnKF that avoid the insertion of $E\k$ altogether. 

Similar to the KF, the gain $\bar K\k$ should be obtained from the solution of a linear system of equations
\begin{equation}
\bar K\k \widetilde Y\kkm \widetilde Y\kkm\T = \widetilde X\kkm \widetilde Y\kkm\T \label{eq:enkfGainLS1}.
\end{equation}



\section{Some Properties and Challenges of the EnKF}
\label{sec:enkfProp}

After a brief review of convergence results and the computational complexity of the EnKF, we discuss adverse effects that can occur in EnKF with finite ensemble size $N$.

\subsection{Asymptotic convergence results}

In linear systems the EnKF mean and covariance~\eqref{eq:sampleStat} converge to the KF results~\eqref{eq:kfMeas} as $N\rightarrow\infty$. This result has been established from different theoretical perspectives~\cite{furrer_estimation_2007,butala_asymptotic_2008, crisan_large_2011, mandel_convergence_2011}. 
%

For nonlinear systems the convergence is not as tangible. An investigation of the EnKF as particle system is given in~\cite{crisan_large_2011}, with the outcome that the EnKF does not give the Bayesian filtering solution except for the linear Gaussian case. An illustration of this property is given in the example of Sec.~\ref{sec:examplesParticle}. 


\subsection{Computational complexity}

For the complexity analysis we assume that we are only interested in the filtering results and that $n>N>m$, that is, the number of measurements is less than the ensemble size and state dimension.

The KF propagates the $n$-dimensional mean vector $\xh\kk$ and the $n\times n$ covariance matrix $P\kk$ with $n(n+1)/2$ unique entries. These storage requirements of $\ordo(n^2/2)$ dominate for large $n>m$. The EnKF requires the storage of only $nN$ values. The space required to store the Kalman gain and other intermediate results is similar for the KF and EnKF. A reduction via sequential processing of measurements, as explained in Sec.~\ref{sec:enkfExtSequential}, is possible for both. 

For large $n$ the computational bottleneck of the KF is the covariance time update~\eqref{eq:kfTimeCov}. Without considering any potential structure in $F$, slightly less than $\ordo(n^3)$ floating point operations (flops) are required. Contemporary matrix multiplication routines~\cite{trefethen_numerical_1997} achieve a reduction to roughly $\ordo(n^{2.4})$. The EnKF time update requires the propagation of $N$ realizations. If each propagation costs $\ordo(n^2)$ flops, then time update is achieved in $\ordo(n^2N)$ flops. 

The computation of the KF gain requires $\ordo(n^2m)$ flops for the computation of $M\k$ and $S\k$. The solution of~\eqref{eq:kfGain} for $K\k$ amounts to $\ordo(m^3)$. The actual measurement update~\eqref{eq:kfMeas} adds further $\ordo(n^2m)$ flops. For large $n$, the total cost is $\ordo(n^2m)$. In contrast, the EnKF parameters $\bar M\k$ and $\bar S\k$ can be computed in $\ordo(nmN)$ flops which, again, dominates the total cost of the measurement update for large $n$. So, the EnKF reduces the flop count by a factor $\frac{N}{n}$.

\subsection{Sampling and coupling effects for finite ensemble size}

A serious issue in the EnKF is a commonly noted tendency to underestimate the state uncertainty when using $N<n$ ensemble members~\cite{houtekamer_data_1998,houtekamer_ensemble_2005,hamill_ensemble-based_2006}. In other words, the EnKF becomes over-confident and is likely to diverge~\cite{anderson_optimal_1979} for too small $N$. A number of causes and related effects can be noted. 

First, an ensemble $X\kkm$ with too few members might not cover the relevant regions of the state-space well enough after the time update~\eqref{eq:enkfPred}. The underestimated spread persists in the measurement update~\eqref{eq:enkfMeas1} or~\eqref{eq:enkfMeas2} and also $X\kk$ shows too little spread. 

Second, the ensemble can only transport limited information and provide a sampling covariance $\bar P\kk$, \eqref{eq:sampleCov} or~\eqref{eq:sampleCov2}, of at most rank $N-1$. Consequently, identically zero entries of $P\kk$ are difficult to reproduce and unwanted spurious correlations show up in $\bar P\kk$. An example would be an unreasonably large correlation between the temperature at two distant locations on the globe. Of course, these correlations also affect $\bar M\k$ and $\bar S\k$, and thus the EnKF gain $\bar K\k$ in \eqref{eq:enkfGainLS1}. As a result, state components that are actually uncorrelated to $y\k$ are erroneously updated in~\eqref{eq:enkfMeas1} or~\eqref{eq:enkfMeas2}. Again, this leads to a reduction in ensemble spread. 

Third, the ensemble members are nonlinearly coupled because the gain~\eqref{eq:enkfGainLS1} is computed from the ensemble. This ``inbreeding''~\cite{houtekamer_data_1998} increases with each measurement update. An interesting side effect is that the ensemble is not independent and Gaussian, even for linear Gaussian problems. To illustrate this, we combine \eqref{eq:enkfGainLS1} and \eqref{eq:enkfMeas2} to obtain
\begin{equation}
X\kk = X\kkm + (\widetilde X\kkm \widetilde Y\kkm\T) (\widetilde Y\kkm \widetilde Y\kkm\T)\inv (y\k\one\T - Y\kkm)
\end{equation}
and consider a linear model~\eqref{eq:ssm} with $n=1$, $H=1$, and a zero-mean $X\kkm$. Then, one member of $X\kk$ is given by
\begin{equation}
x\kk\sui = x\kkm\sui + \frac{\sum_{j=1}^N (x\kkm\suj)^2}{\sum_{j=1}^N (x\kkm\suj + e\k\suj)^2} (y\k - x\kkm\sui - e\kkm\sui), \label{eq:ensMemb}
\end{equation}
which clearly shows the nonlinear dependencies that impede Gaussianity of $x\kk\sui$. Although similar conclusions hold for the general case, concise effects on the ensemble spread are difficult to analyze. Some special cases ($n=1$ and $n=m$, $H=I$, $R\propto I$) are investigated in~\cite{furrer_estimation_2007} and shown to produce an underestimated $\bar P\kk$.

Finally, the random sampling in the measurement update by inserting measurement noise in~\eqref{eq:enkfPredMeas} or~\eqref{eq:enkfPredMeasNonlin} adds to the EnKF error budget. The inherent sampling errors can be reduced by using the square root EnKF of Sec.~\ref{sec:enkfExtSquareRoot}. 


Experiments suggest that there is a threshold for $N$ above which the EnKF works. A good example is given in~\cite{ott_local_2004}. Sec.~\ref{sec:enkfExt} discusses methods such as inflation and localization that can reduce this minimum $N$. 

\section{Important Extensions to the EnKF}
\label{sec:enkfExt}

The previous section highlighted some of the challenges of the EnKF. Here, we summarize the important extensions that are often essential to achieve a working EnKF with only few ensemble members. 

\subsection{Sequential updates}
\label{sec:enkfExtSequential}

For the KF it is algebraically equivalent to carry out $m$ measurement updates~\eqref{eq:kfMeas} with the scalar components of $y\k$ instead of a batch update with the $m$-dimensional $y\k$, if the measurement noise covariance $R$ is diagonal~\cite{anderson_optimal_1979}. Although often treated as a side note only, this technique is very useful. It yields a more flexible algorithm with regard to the availability of measurements at each time step $k$ and reduces the computational complexity. After all, the Kalman gain~\eqref{eq:kfGain} merely requires a scalar division for each component of $y\k$. An extension to block-diagonal $R$ is imminent.

Motivated by the large number of measurements in geoscientific problems, sequential updates have also been suggested for the EnKF~\cite{houtekamer_sequential_2001}. Because of the randomness inherent to the EnKF, there is no algebraic equivalence between sequential and batch updates. Hence, the order in which measurements are processed has an effect on the filtering results. 

Furthermore, an unusual alternative interpretation of sequential updates can be found in the EnKF literature. Namely, measurement updates are carried out ``grid point by grid point''~\cite{houtekamer_data_1998,evensen_ensemble_2003,ott_local_2004}, that is, an iteration is carried out over state rather than measurement components. We will return to this aspect in Sec.~\ref{sec:enkfExtLocalization}.

\subsection{Model knowledge in the EnKF and square-root filters}\label{sec:enkfExtSquareRoot}

The sampling based derivation of the EnKF in equations~\eqref{eq:enkfPred} through~\eqref{eq:enkfGainLS1} facilitates a compact presentation. However, the randomization through $E\k$ in~\eqref{eq:enkfPredMeas} or~\eqref{eq:enkfPredMeasNonlin} adds Monte Carlo sampling errors to the EnKF budget. This section discusses how these errors can be reduced for linear systems~\eqref{eq:ssm}. Similar results for nonlinear systems with additive noise follow easily. The interpretation of ensembles as (rectangular) matrix square roots is a common theme in the following approaches. In~\eqref{eq:sampleCov2}, for instance, $\tfrac{1}{\sqrt{N-1}}\widetilde X\kk$ can be seen as an $n\times N$ square root of $\bar P\kk$.  


A first thing to note is that the cross covariance $M\k$ in the KF and its ensemble equivalent $\bar M\k$ should not be influenced by additive measurement noise $e\k$. Therefore, it is reasonable to replace $\widetilde Y\kkm$ with 
\begin{subequations}\label{eq:msModel}
\begin{equation}
\widetilde Z\kkm = H \widetilde X\kkm \label{eq:zEns}
\end{equation}
so as to reduce the Monte Carlo variance of~\eqref{eq:msSamp} using
\begin{align}
\bar M\k &= \tfrac{1}{N-1}\widetilde X\kkm \widetilde Z\kkm\T,\\
\bar S\k &= \tfrac{1}{N-1}\widetilde Z\kkm \widetilde Z\kkm\T + R. \label{eq:msModelS}
\end{align}
\end{subequations}
%
%
The Kalman gain $\bar K\k$ is then computed as in the KF~\eqref{eq:kfGain}. Alternatively, a matrix square-root $R\sr$ with $R\sr R\srt=R$ can be used to factorize
\begin{equation}
\bar S\k = \bbm \tfrac{1}{\sqrt{N-1}}\widetilde Z\kkm & R\sr \ebm  \bbm \tfrac{1}{\sqrt{N-1}}\widetilde Z\kkm\T \\ R\srt \ebm.
\end{equation}
A QR decomposition~\cite{trefethen_numerical_1997} of the right matrix then yields a triangular $m\times m$ square root of $\bar S\k$, and the computation of $\bar K\k$ simplifies to forward and backward substitution. Such ideas have their origin in sigma point KF variants~\cite{norgaard_new_2000}. 

The KF permits offline computation of the covariance matrices $P\kk$ for all $k$ because they do not depend on the measurements. In an EnKF for a linear system~\eqref{eq:ssm} we can mimic this behavior by propagating zero-mean ensembles $\widetilde X\kk$ that only carry the information of $P\kk$. This is the central idea of different square root EnKF~\cite{tippett_ensemble_2003} which were suggested in~\cite{anderson_ensemble_2001,bishop_adaptive_2001,whitaker_ensemble_2002}. The name stems from a relation to square root KF~\cite{anderson_optimal_1979} which propagate $n\times n$ matrix square roots $\smash{P\sr\kk}$ with $\smash{P\sr\kk P\srt\kk=P\kk}$. Most importantly, the artificial measurement noise and the inherent sampling error can be avoided. 

The following derivation~\cite{tippett_ensemble_2003} rewrites an alternative expression for~\eqref{eq:kfMeasCov} using a square root $\smash{P\sr\kkm}$ and its ensemble approximation $\tfrac{1}{N-1} \widetilde X\kkm$:
\begin{subequations}
\begin{align}
\nonumber P\kk &= (I-K\k H)P\kkm\\
&= P\sr\kkm (I-P\srt\kkm H\T S\k\inv H P\sr\kkm) P\srt\kkm\\
\nonumber &\approx \tfrac{1}{N-1} \widetilde X\kkm (I-\tfrac{1}{N-1}\widetilde Z\kkm\T \bar S\k\inv \widetilde Z\kkm) \widetilde X\kkm\T,
\end{align}
where~\eqref{eq:zEns} was used. 
The next step is to factorize
\begin{equation}\label{eq:piFac}
(I-\tfrac{1}{N-1}\widetilde Z\kkm\T \bar S\k\inv \widetilde Z\kkm) = \Pi\k\sr\Pi\k\srt,
\end{equation}
\end{subequations}
which requires the left hand side to be positive definite. This property is easily established for the positive definite $\bar S\k$ of~\eqref{eq:msModelS} after realizing that the left hand side of~\eqref{eq:piFac} is a Schur complement~\cite{trefethen_numerical_1997} of a positive definite matrix.

Finally, the $N\times N$ matrix $\Pi\k\sr$ can be used to create a deviation ensemble
\begin{equation}
\widetilde X\kk = \widetilde X\kkm \Pi\k\sr
\end{equation}
that correctly encodes $P\kk$ without using any random perturbations. Other variants update the deviation ensemble via a multiplication from the left~\cite{anderson_ensemble_2001}, which is more costly for large $n$. Some more conditions on $\Pi\k\sr$ must be met for $\widetilde X\kk$ to remain zero-mean~\cite{sakov_implications_2008,livings_unbiased_2008}. 

The actual filtering is achieved by updating a single estimate according to
\begin{equation}
\bar x\kk = (I-\bar K\k H) F \bar x\kmkm + \bar K\k y\k,
\end{equation}
where $\bar K\k$ is computed from the deviation ensembles. 


There are indications that in nonlinear and non-Gaussian systems the sampling based EnKF variants should be preferable over their square root counterparts: A low-dimensional example is studied in \cite{lawson_implications_2004}; the impression is confirmed for a high-dimensional problem in~\cite{leeuwenburgh_impact_2005}.

\subsection{Ensemble inflation}

Ensemble or covariance inflation is a measure to counteract the tendency of the EnKF to underestimate the state uncertainty for small $N$, and an important ingredient in operational EnKF~\cite{hamill_ensemble-based_2006}. The spread of the prediction ensemble~$X\kkm$ is increased according to
\begin{equation}
X\kkm = \bar x\kkm\one\T + c \widetilde X\kkm
\end{equation}
with a factor $c>1$. In the EnKF context, this heuristic has been proposed in~\cite{anderson_monte_1999}. Related concepts are dithering in the PF~\cite{gustafsson_particle_2010} and the ``fudge factor'' to increase $P\kkm$ in the KF~\cite{bar-shalom_estimation_2001}.
Extensions to adaptive inflation, where $c$ is adjusted online, are discussed in~\cite{anderson_ensemble_2009}.

\subsection{Localization}
\label{sec:enkfExtLocalization}

Localization is a technique to address the issue of spurious correlations in the EnKF, and a crucial feature of operational EnKF~\cite{hamill_ensemble-based_2006,houtekamer_ensemble_2005}. The underlying idea applies equally well to the EnKF and the KF, and can be used to systematically update only a part of the state vector with each measurement. 

In order to explain the concept, we regard the KF measurement update for a linear system~\eqref{eq:ssm} with a low-dimensional\footnotemark\ measurement $y\k$. 
\footnotetext{We assume that the components can be processed sequentially.}  
Let $x=x\kkm$ and $P=P\kkm$ for notational convenience. It is possible to permute the state components such that 
\begin{equation}\label{eq:locStructure}
x = \bbm x_1\\ x_2\\ x_3 \ebm, \quad H = \bbm H_1 & 0 & 0 \ebm, \quad P = \bbm P_1 & P_{12} & 0\\ P_{12}\T & P_2 & P_{23}\T\\ 0 & P_{23}\T & P_{3}\ebm.
\end{equation}
Only the part $x_1$ appears in the measurement equation~\eqref{eq:meas} $y\k = H_1 x_1 + e\k$. While $x_2$ is correlated to $x_1$, there is zero correlation between $x_1$ and $x_3$. As a consequence, many submatrices of $P$ vanish in the computation of
\begin{subequations}\label{eq:locStructureGain}
\begin{align}
PH\T &= \bbm H_1 P_1 & H_1 P_{12} & 0 \ebm\T,\\
HPH\T &= H_1 P_1 H_1\T,
\end{align}
and do not contribute to the Kalman gain~\eqref{eq:kfGain}
\begin{equation}
K\k = \bbm P_1 H_1\T \\ P_{12}\T H_1\T \\0\ebm (H_1 P_1 H_1\T + R)\inv.
\end{equation}
\end{subequations}
A KF measurement update~\eqref{eq:kfMeas} with the above $K\k$ does not affect the $x_3$ estimate or covariance. Hence, there is a lower-dimensional measurement update that only alters the statistics of $x_1$ and $x_2$.

Localization in the EnKF enforces the above structure using two prevailing techniques, local updates~\cite{houtekamer_data_1998,evensen_ensemble_2003,ott_local_2004} and covariance tapering~\cite{houtekamer_sequential_2001,hamill_distance-dependent_2001}. Both rely on prior knowledge of the covariance structure. For example, the state components are often connected to geographic locations in geoscientific applications. From the underlying physics it is reasonable to assume zero correlation between distant states. Unfortunately, this viewpoint is not transferable to high-dimensional problems in general. 

Local updates were introduced for the sampling based EnKF in~\cite{houtekamer_data_1998} and for different square root EnKF in~\cite{evensen_ensemble_2003,ott_local_2004}. Nonlinear measurement functions~\eqref{eq:measNonlin} are linearized in the latter two. All of the above references update the state vector ``grid point by grid point'', which appears unusual from a KF perspective~\cite{anderson_optimal_1979}. In an iteration, local state vectors of small dimension ($<N$) are chosen and updated with a subset of supposedly relevant measurements. These ``full rank'' updates avoid some of the problems associated with small $N$ and large $n$. However, discontinuities between state components are introduced~\cite{sakov_relation_2010}. Some heuristics to combine the local ensembles and further implementation details can be found in~\cite{ott_local_2004,hunt_efficient_2007}. 

Under the assumption of the structure in~\eqref{eq:locStructure}, a local analysis would amount to an EnKF update of the $x_1$- and $x_2$-components only, to avoid errors in $x_3$. 


Covariance tapering was introduced in~\cite{houtekamer_data_1998}. It contradicts the EnKF idea in the sense that the ensemble covariance $\bar P\kkm$ of $X\kkm$ is processed. However, it will become clear that not all entries of $\bar P\kkm$ must be computed. Prior knowledge of a covariance structure as in~\eqref{eq:locStructure} is used to create an $n\times n$ matrix $\rho$ with entries in $[0,1]$, and a tapered covariance $(\rho \circ \bar P\kkm)$
%
%
is computed. Here, $\circ$ denotes the element-wise Hadamard or Schur product~\cite{trefethen_numerical_1997}. A typical $\rho$ has ones on the diagonal and decays smoothly to zero for unwanted off-diagonal elements. The standard choice uses a compactly supported correlation function from~\cite{gaspari_construction_1999} and is discussed in~\cite{houtekamer_sequential_2001,hamill_distance-dependent_2001,sakov_relation_2010}. Subsequently, the Kalman gain is computed as in the KF~\eqref{eq:kfGain} using
\begin{subequations}\label{eq:taperedMS}
\begin{align}
\bar M\k &= (\rho \circ \bar P\kkm) H\T, \label{eq:taperedM}\\
\bar S\k &= H(\rho \circ \bar P\kkm) H\T + R,
\end{align}
\end{subequations}
where we assumed a linear measurement relation~\eqref{eq:meas}.

There are some technicalities associated with the tapering operation. Only positive semi-definite $\rho$ guarantee that $(\rho \circ \bar P\kkm)$ is a valid covariance~\cite{furrer_estimation_2007}. Full rank $\rho$ yield an increased rank in $(\rho \circ \bar P\kkm)$~\cite{houtekamer_sequential_2001}. However, low rank $\rho$ do not necessarily decrease the rank of $(\rho \circ \bar P\kkm)$. A closely related problem to finding valid (positive semi-definite or definite) $\rho$ is the creation of covariance functions and kernels in Gaussian processes~\cite{rasmussen_gaussian_2005}. Here, a methodology to create more complicated kernels from simpler ones could be used to create $\rho$. 

Unfortunately, the Hadamard product cannot be formulated as an operation on the ensembles in general. Still, the computational requirements can be limited by only working with the non-zero elements of $(\rho \circ \bar P\kkm)$. Furthermore, it is common to avoid the computation of $\bar P\kkm$ using
%
%
\begin{equation}
\bar M\k = \rho_M \circ \bar M\k, \label{eq:taperedM2}
\end{equation}
instead of~\eqref{eq:taperedM} and to skip the tapering in $S\k$ altogether~\cite{hamill_distance-dependent_2001}. After all, for low-dimensional $y\k$ (small $m$) $\bar M\k$ has the strongest influence on the gain $\bar K\k$. Also the matrix $\rho_M$ is constructed from prior knowledge about the correlation. In the geoscientific context, where the state components and measurements are associated with geographic locations, this is easy. In general, however, it might not be possible to devise an appropriate $\rho_M$. Other variants~\cite{houtekamer_sequential_2001,furrer_estimation_2007,sakov_relation_2010} with tapering for $\bar S\k$ exist and have in common that they are only identical to~\eqref{eq:taperedMS} for $H=I$. 

Some relations between local updates and covariance tapering are discussed in~\cite{sakov_relation_2010}. For the structure in~\eqref{eq:locStructure} we can suggest a {rank-1} taper $\rho$ that establishes a correspondence between the two concepts. Let $r_1$ and $r_2$ be vectors of the same dimensions as $x_1$ and $x_2$, respectively, that contain all ones. Let $r_3$ be a zero vector of the same dimension as $x_3$ and $r\T=[r_1\T, r_2\T, r_3\T]$. Then $\rho = rr\T$ removes all entries from $\bar P\kkm$ that would disappear in~\eqref{eq:locStructureGain} anyhow. Furthermore, the Hadamard product for the rank-1 $\rho$ can be written as an operation on the ensemble $\widetilde X\kkm$ using
%
\begin{align}
\nonumber (rr\T)\circ \bar P\kkm
&= \diag(r) \bar P\kkm \diag(r)\\
&= \tfrac{1}{N-1} \left(\diag(r) \widetilde X\kkm\right)\left(\diag(r) \widetilde X\kkm  \right)\T.
\end{align}
The multiplication with $\diag(r)$ merely removes the rows corresponding to $x_3$, which establishes an equivalence between local updates and covariance tapering. By picking a smoothly decaying $r$ we can furthermore avoid the discontinuities associated with local updates. 


\subsection{The EnKF gain and least squares}

A parallel to least squares problems can be disclosed by closer inspection of the equation~\eqref{eq:enkfGainLS1} that is used to compute the EnKF gain $\bar K\k$. Perhaps more apparent in the transpose of~\eqref{eq:enkfGainLS1}, in
\begin{subequations}
\begin{equation}
\widetilde Y\kkm \widetilde Y\kkm\T \bar K\k\T  = \widetilde Y\kkm \widetilde X\kkm\T, \label{eq:enkfGainLS2}
\end{equation}
appear the normal equations of the least squares problems 
\begin{equation}
\widetilde Y\kkm\T \bar K\k\T = \widetilde X\kkm\T \label{eq:enkfGainLS3}
\end{equation}
\end{subequations}
that are to be solved for each row of $\bar K\k$ and $\widetilde X\kkm$.
Hence, the EnKF iteration can be carried out without explicitly computing any sample covariance matrices if instead efficient solutions to the problem~\eqref{eq:enkfGainLS3} are employed.
Furthermore, the problem~\eqref{eq:enkfGainLS3} could be modified using regularization~\cite{hastie_elements_2011} to enforce sparsity in $\bar K\k$. This would be an alternative approach to the localization methods discussed earlier. Related ideas to improve the Kalman gain using bootstrap methods~\cite{hastie_elements_2011} for computing $\bar M\k$ and $\bar S\k$ in~\eqref{eq:msSamp} are discussed in~\cite{zhang_improving_2010,myrseth_resampling_2013}.


\section{Relations to other algorithms}
\label{sec:enkfRelations}

The EnKF for nonlinear systems~\eqref{eq:ssmNonlin} differs from other sampling based nonlinear filters such as sigma point KF~\cite{roth_nonlinear_2016} or particle filters (PF)~\cite{gustafsson_particle_2010}.
One reason for this is that the EnKF approximates the KF algorithm (with the side effect that it can be applied to \eqref{eq:ssmNonlin}) rather than trying to solve the nonlinear filtering problem directly. 

The biggest difference between the EnKF and sigma point filters~\cite{roth_nonlinear_2016} such as the unscented KF~\cite{julier_new_1995,julier_unscented_2004} or divided difference KF~\cite{norgaard_new_2000} is the measurement update. Whereas the EnKF updates its ensembles, the latter carry out the KF measurement update~\eqref{eq:kfMeas} using approximately computed mean values and covariance matrices. That is, the samples or sigma points are condensed into a filtering estimate $\xh\kk$ and its covariance $P\kk$, which entails a loss of information and can be seen as an inherent Gaussian assumption on the filtering density $p(x\k|y\otk)$. In contrast, the EnKF can preserve more information and deviations from Gaussianity in the ensemble. Similarities appear in the gain computations of the EnKF and sigma point KF. In both, the Kalman gain appears as a function of the sampling covariance matrices, although with the deterministic sigma points and weights in the latter. With their origin in the KF, both sigma point filters and the EnKF can be expected to share difficulties with multimodal posterior distributions. 

Similar to the EnKF, the PF propagates $N$ state realizations that are called particles. For the bootstrap particle filter~\cite{gordon_novel_1993}, the prediction step corresponds to the EnKF time update~\eqref{eq:enkfPredNonlin}. Apart from that, however, the differences dominate. First, the PF is designed as an approximate solution of the Bayesian filtering equations~\cite{jazwinski_stochastic_1970} using sequential importance sampling~\cite{gustafsson_particle_2010}. For $N\rightarrow\infty$, the PF solution recovers the true filtering density. 
Second, the samples in basic PF variants are generated from a proposal distribution only once every time instance and then left untouched. The measurement update amounts to updating the particle weights, which leads to a degeneracy problem for large $n$. In the EnKF, in contrast, the ensemble members are influenced by the time and the measurement update. Third, the PF relies on a crucial resampling step that is not present in the EnKF.

In summary, the EnKF appears as a distinct algorithm besides sigma point KF and PF. Its properties and potential for nonlinear problems remain to be fully investigated. Existing results that the EnKF does not converge to the Bayesian filtering recursion~\cite{crisan_large_2011} remain to be interpreted in a constructive manner. 

\section{Instructive Simulation Examples}
\label{sec:examples}

Four examples are discussed in greater detail, among them one popular benchmark problem of the SP and DA literature each. 


\subsection{A scalar linear Gaussian model}
\label{sec:examples_1d}
 

The first example illustrates the tendency of the EnKF to underestimate the state uncertainty. A related example is studied in~\cite{whitaker_ensemble_2002}. We compare the EnKF variance $\bar P\kk$ to the $P\kk$ of the KF via Monte Carlo simulations on the simple scalar state-space model
\begin{subequations}\label{eq:exScalarModel}
\begin{align}
x\kp &= x\k + v\k,\\
y\k &= x\k + e\k.
\end{align}
%
The initial state $x_0$, the process noise $v_k$, and the measurement noise $e\k$ are specified by the probability density functions
\begin{align}
p(x_0) &= \N(x_0;0,0.1),\\
p(v\k) &= \N(v\k;0,0.1),\\
p(e\k) &= \N(e\k;0,0.01).
\end{align}
\end{subequations}
A trajectory of~\eqref{eq:exScalarModel} is simulated and a KF is used to compute the optimal variances $P\kk$. Because the model is time-invariant, the $P\kk$ quickly converge to a constant value. For $k>3$ $P\kk=0.0092$ is obtained.

\begin{figure}
 \centering
 \includegraphics[]{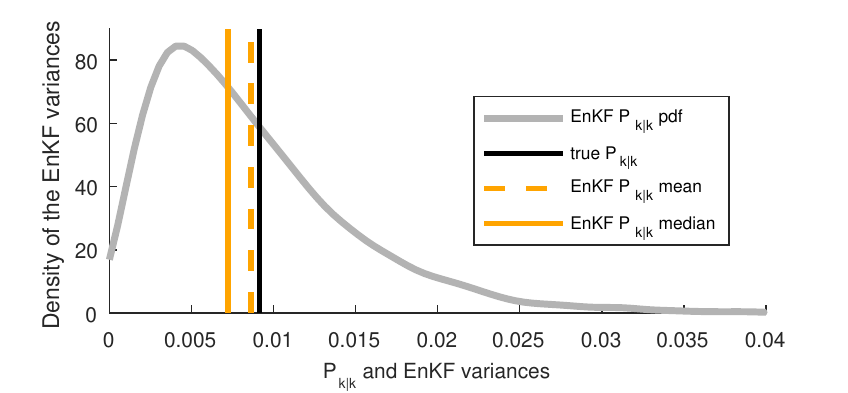}
 \caption{Distribution of EnKF variances $\bar P\kk$ with $k=10$ and $N=5$ ensemble members for $10000$ runs on the same trajectory. Also shown is the mean and median of all outcomes and the desired KF variance $P\kk$.}
 \label{fig:pHist1D_k10_N5_nMc10000_enkfGain}
\end{figure}

Next, 10000 Monte Carlo experiments with a sampling based EnKF with $N=5$ are performed. The distribution of obtained $\bar P\kk$ for $k=10$ is illustrated in Fig.~\ref{fig:pHist1D_k10_N5_nMc10000_enkfGain}. The vertical lines indicate the $P\kk$ of the KF and the median and mean of the $\bar P\kk$ outcomes.
The average $\bar P\kk$ over the Monte Carlo realizations is close to the desired $P\kk$. However, there is a large spread among the~$\bar P\kk$ and the distribution is skewed toward zero with its median below $P\kk$. Although $N>n$, there is a tendency to underestimate $P\kk$.

\begin{figure}
 \centering
 \includegraphics[]{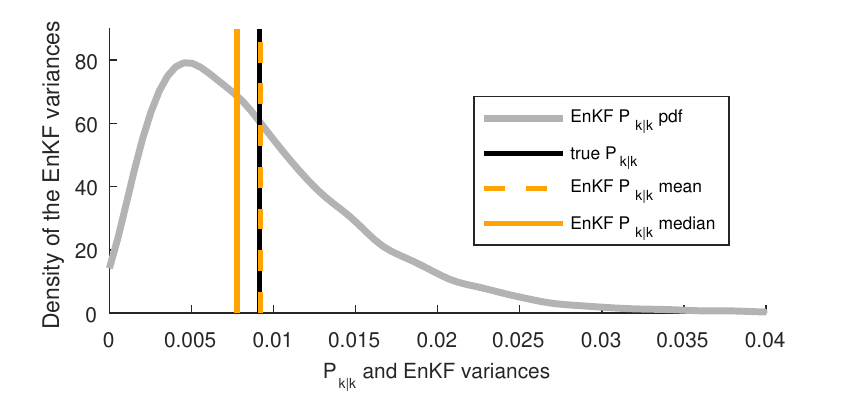}
 \caption{Distribution of EnKF variances $\bar P\kk$, but computed with the correct Kalman gain. Otherwise, similar to Fig.~\ref{fig:pHist1D_k10_N5_nMc10000_enkfGain}.}
 \label{fig:pHist1D_k10_N5_nMc10000_statGain}
\end{figure}

In order to clarify the reason for this behavior and whether it has to do with the coupling between the EnKF $\bar K\k$ and the ensemble members, we repeat the experiment with an EnKF that uses the gain of the stationary KF for all $k$. The resulting outcomes are illustrated in Fig.~\ref{fig:pHist1D_k10_N5_nMc10000_statGain}.
%
Now, the average $\bar P\kk$ is correct. However, the median shows that there is still more probability mass below $P\kk$. The tendency to underestimate $P\kk$ and the remaining spread must be due to random sampling errors.
For larger $N$ the effect vanishes, and the median and mean of $\bar P\kk$ appear similar for $N\geq 10$.


\subsection{The particle filter benchmark}
\label{sec:examplesParticle}


In the second example we show that the EnKF does not converge to the Bayesian filtering solution in nonlinear systems as $N\rightarrow\infty$~\cite{crisan_large_2011}. A well-known benchmark problem from the PF literature~\cite{gordon_novel_1993} is used. The model is specified by
\begin{subequations}
\begin{align}
x\kp &= \frac{x\k}{2} + 25\frac{x\k}{1+x\k^2} + 8\cos(1.2(k+1)) + v\k, \label{eq:exPfState}\\
y\k &= \frac{x\k^2}{20} + e\k, \label{eq:exPfMeas}
\end{align}
\end{subequations}
with independent $v\k\sim\N(0,10)$, $e\k\sim\N(0,1)$, and $x_0\sim\N(0,1)$. Because the model is scalar, the Bayesian filtering densities $p(x\k\gvn y\otk)$ can be computed numerically using point mass filters (PMF)~\cite{roth_computation_2017}. A sampling based EnKF with $N=500$ is tested and kernel density estimates are used to obtain approximations of $p(x\k\gvn y\otk)$ from the ensembles. For comparison, we include a closely related sigma point KF variant that uses Monte Carlo integration with $N=500$ samples~\cite{roth_nonlinear_2016}. The only difference to the EnKF is that this Monte Carlo KF (MCKF) carries out the KF measurement update~\eqref{eq:kfMeas} to propagate a mean and a variance. We illustrate the results as Gaussian densities.

\begin{figure}
 \centering
 \includegraphics[]{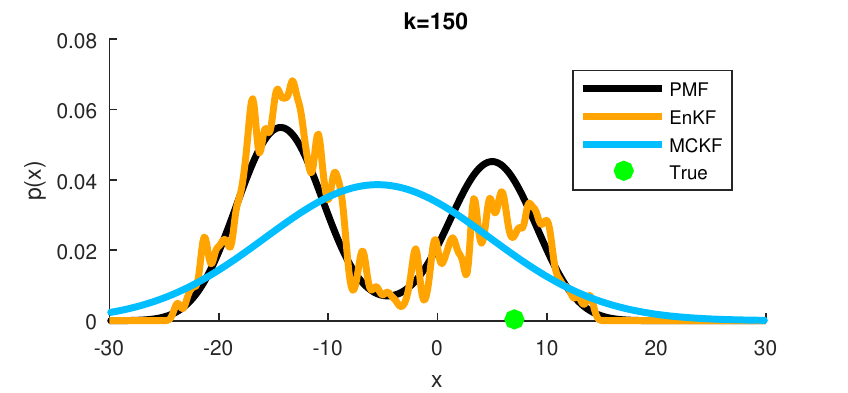}
 \caption{Prediction densities $p(x\k\gvn y\otkm)$ by the PMF, EnKF, and MCKF for $k=150$. The true state is illustrated as green dot. The PMF serves as reference solution.}
 \label{fig:growthEnkfPred}
\end{figure}

Fig.~\ref{fig:growthEnkfPred} shows the prediction results for $k=150$. The PMF reference solution is bimodal with one mode close to the true state. The reason for this lies in the squared $x\k$ in~\eqref{eq:exPfMeas}.
The EnKF prediction resembles the PMF well except for the random variations in the kernel density estimate. The MCKF cannot represent the multimodality but the Gaussian bell covers the relevant regions.

\begin{figure}
 \centering
 \includegraphics[]{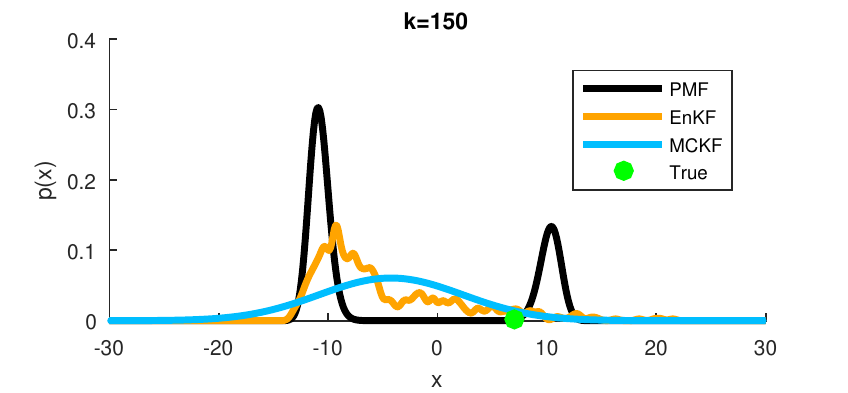}
 \caption{Filtering densities $p(x\k\gvn y\otk)$ by PMF, EnKF, and MCKF for $k=150$. Otherwise similar to~Fig.~\ref{fig:growthEnkfPred}.}
 \label{fig:growthEnkfFilt}
\end{figure}

The filtering results for $k=150$ are shown in Fig.~\ref{fig:growthEnkfFilt}. 
The PMF reference solution has much narrower peaks after including $y\k$. 
The EnKF provides a skewed density that does not resemble $p(x\k\gvn y\otk)$ even though the EnKF prediction approximated $p(x\k\gvn y\otkm)$ well. This is the main take-away result and confirms~\cite{crisan_large_2011}. Again, the MCKF exhibits a large variance.

\begin{figure}
 \centering
 \includegraphics[]{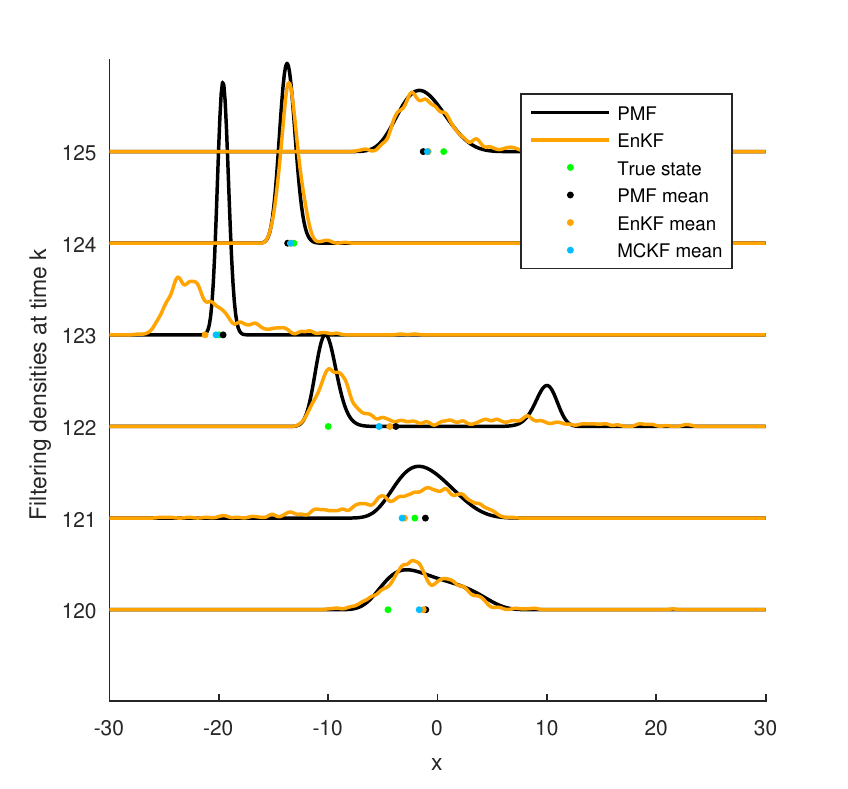}
 \caption{Consecutive filtering densities $p(x\k\gvn y\otk)$ by PMF, EnKF, and MCKF for $k=120,\dotsc,125$. Also illustrated are the mean values of the respective densities and the true state.}
 \label{fig:growthEnkfFiltMore}
\end{figure}

\begin{figure}
 \centering
 \includegraphics[]{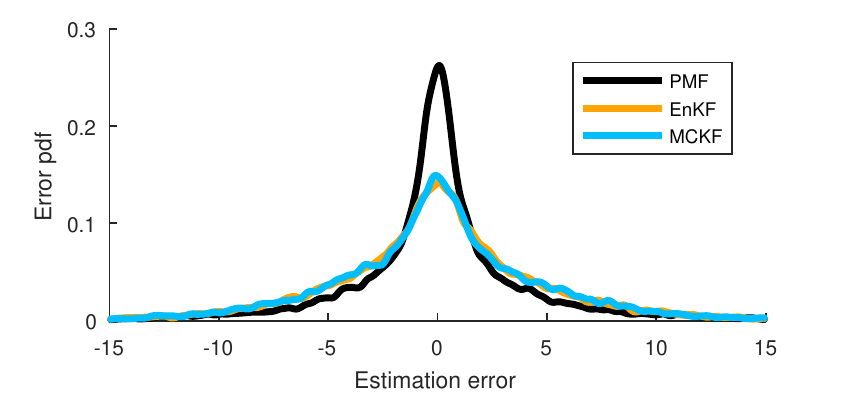}
 \caption{Density of the estimation errors obtained from $100$ Monte Carlo runs with $151$ time steps each.}
 \label{fig:growthEnkfError}
\end{figure}

Further filtering results for the PMF and EnKF are shown in Fig.~\ref{fig:growthEnkfFiltMore}.
It can be seen that the EnKF solutions sometimes resemble the PMF very well, but not always. Similar statements can be made for the prediction results. Dots in~Fig.~\ref{fig:growthEnkfFiltMore} illustrate the mean values as state estimates. Especially for the PMF, it can be seen that the mean (though optimal in a minimum variance sense~\cite{anderson_optimal_1979}) is debatable for multimodal densities. Often, all estimates are quite close. Fig.~\ref{fig:growthEnkfError} provides the estimation error densities obtained from $100$ Monte Carlo experiments with $151$ time steps each. 
The PMF mean estimates exhibit a larger peak around $0$. The estimation errors for the EnKF and MCKF appear similar. This is surprising because the latter employs a Gaussian approximation at each time step. Both error densities have heavier tails than the PMF density. All estimation errors appear unbiased.

\subsection{Batch smoothing using the EnKF}
\label{sec:examples_batch}

We here show how to use the EnKF as smoothing algorithm by estimating batches of states. This allows us to compare its performance for $N< n$ in problems of arbitrary dimension. 

First, we formulate an ``augmented state'' that comprises an entire trajectory of $L+1$ steps,
\begin{equation}
\label{eq:batchState}
\xi = \bbm x_0\T & x_1\T & \dotso & x_L\T \ebm\T,
\end{equation}
with dimension $n=(L+1)n_x$. Second, we note that the measurements $y\k$, $k=1,\dotsc,L$, have uncorrelated measurement noise and known relations to the components of~$\xi$. For linear systems~\eqref{eq:ssm}, the predicted mean and covariance of~$\xi$ can be easily derived, and smoothed estimates of all $x\k$, $k=0,\dotsc,L$, can be obtained by sequentially processing all $y\k$ in KF measurement updates for $\xi$.

Also other smoothing variants and the Rauch-Tung-Striebel (RTS) algorithm can be derived from state augmentation approaches~\cite{anderson_optimal_1979}. Due to its sequential nature, however, the RTS smoother does not provide joint covariance matrices of $x\k$ and $x_{k+i}$ for $i\neq 0$. Except for this and the higher computational complexity of working with $\xi$, the batch and RTS smoothers are equivalent for~\eqref{eq:ssm}. 

An EnKF approach to batch smoothing mimics the above. A prediction ensemble for $\xi$ is obtained by simulating $N$ trajectories for random process noise and initial state realizations. This can also be carried out for nonlinear models~\eqref{eq:ssmNonlin}. Then, sequential EnKF measurement updates are performed for all~$y\k$. 

For our experiments we use a tracking problem with a constant velocity model~\cite{bar-shalom_estimation_2001} and position measurements. The low-dimensional state is given by
\begin{subequations}
\begin{equation}
  x
  = \bbm \mathsf{x} & \mathsf{y} & \dot{\mathsf{x}} & \dot{\mathsf{y}} \ebm\T
\end{equation}
and comprises the Cartesian position [m] and velocity [m/s] of an object. The parameters of~\eqref{eq:ssm} are given by
\begin{align}
  F
  &= \bbm I_2 & \mathsf{T}I_2\\ 0 & I_2 \ebm ,
  &
  G
  &= \bbm \frac{\mathsf{T}^2}{2} I_2 \\ \mathsf{T} I_2 \ebm ,
  &
  H
  &= \bbm I_2 & 0 \ebm,
\end{align}
with $\mathsf{T}=1\,$s. The initial state $x_0$ is Gaussian distributed with
%
\begin{equation}
\xh_0 = \bbm 0& 0& 15& -10\ebm\T,\quad 
P_0 = \diag(50^2,50^2,20^2,20^2),
\end{equation}
and the process and measurement noise covariances are
\begin{align}
Q = \diag(10,50), \quad R=\bbm 2000&1000\\1000&1980 \ebm.
\end{align}
\end{subequations}
With $n_x=4$ and $L=49$ we obtain $n=200$ as dimension of~$\xi$. The RTS solution is compared to EnKF of ensemble size $N=\{10,20,50\}$. Monte Carlo errors are reduced using~\eqref{eq:msModel} in the gain computations.

\begin{figure}
  \centering
  \includegraphics{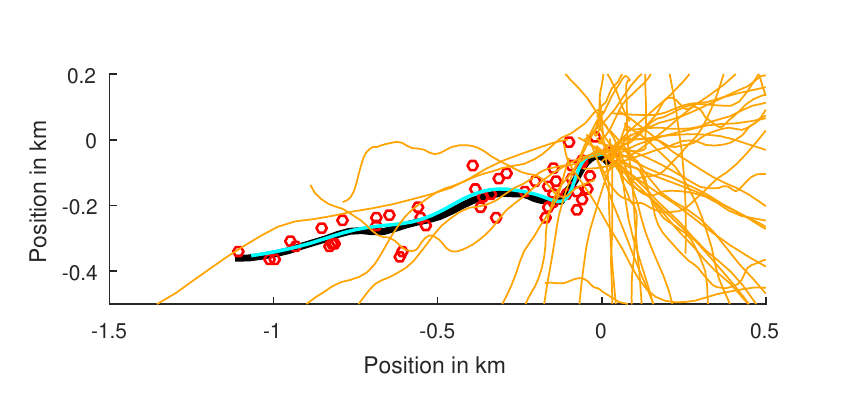}
  \caption{Illustration of a representative trajectory (black), the RTS smoothing solution (cyan), and an initial ensemble ($N=50$, orange). Red circles depict the measurements. Most ensemble trajectories go beyond the plot area.}
  \label{fig:batch_orderforward_N50_meas0_locdefaultInf}
\end{figure}

\begin{figure}
  \centering
  \includegraphics{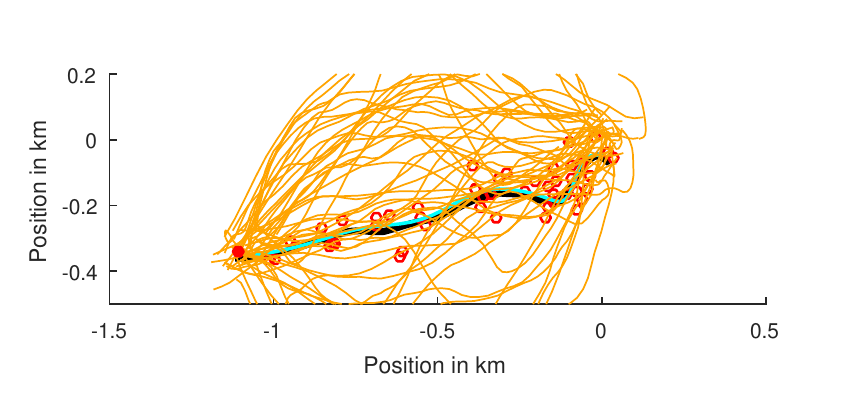}
  \caption{The ensemble of Fig.~\ref{fig:batch_orderforward_N50_meas0_locdefaultInf} after a measurement update with $y_L$ only. Some ensemble trajectories leave and re-enter the plot area.}
  \label{fig:batch_orderreverse_N50_meas1_locdefaultInf}
\end{figure}

\begin{figure}
  \centering
  \includegraphics{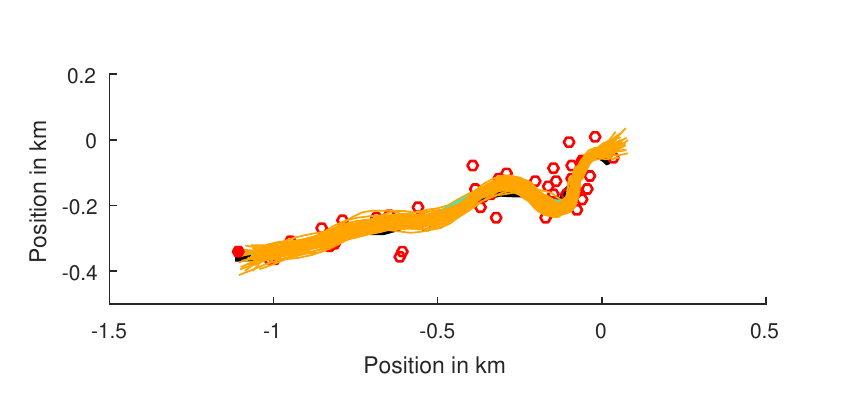}
  \caption{The ensemble of Fig.~\ref{fig:batch_orderforward_N50_meas0_locdefaultInf} after updating with all measurements in the order $y_1, \dotsc y_L$. The RTS solution is covered well.}
  \label{fig:batch_orderforward_N50_meas49_locdefaultInf}
\end{figure}

\begin{figure}
  \centering
  \includegraphics{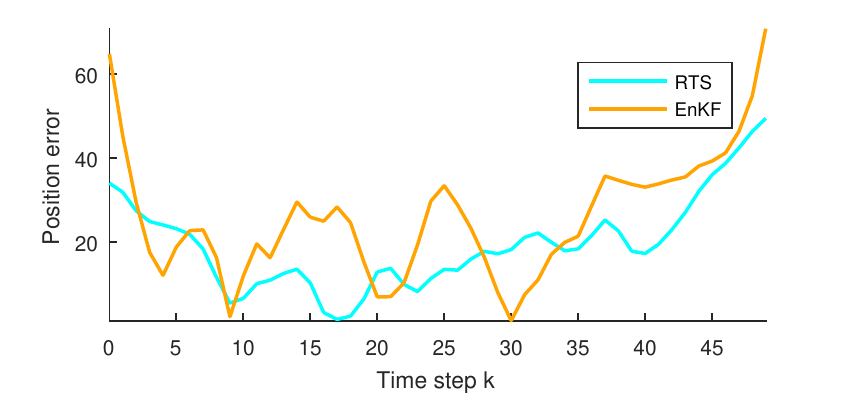}
  \caption{Position errors of the RTS (cyan) and the EnKF ($N=50$, orange) after updating with all measurements in the order $y_1, \dotsc y_L$.}
  \label{fig:batchRMSE_orderforward_N50_locdefaultInf}
\end{figure}

A realization of a true trajectory and its measurements is provided in Fig.~\ref{fig:batch_orderforward_N50_meas0_locdefaultInf} together with the RTS estimate and an ensemble of $N=50$ trajectories. 
The latter are the initial ensemble of an EnKF. The ensemble is well gathered around the initial position but fans out wildly. 
Fig.~\ref{fig:batch_orderreverse_N50_meas1_locdefaultInf} shows the ensemble after an update with $y_L$ only. 
%
The measurement at the end of the trajectory provides an anchor point and quickly reduces the spread of the ensemble. Fig.~\ref{fig:batch_orderforward_N50_meas49_locdefaultInf} shows the result after processing all measurements in sequential order from first to last. 
The true trajectory and the RTS estimate are mostly covered well by the ensemble. The EnKF with $N=50$ appears consistent in this respect. 
Position errors for the RTS and the EnKF are provided in Fig.~\ref{fig:batchRMSE_orderforward_N50_locdefaultInf}. 
%
The EnKF performs slightly worse than the RTS but still gives good results for $N=50$, without extra inflation or localization. 

\begin{figure}
  \centering
  \includegraphics{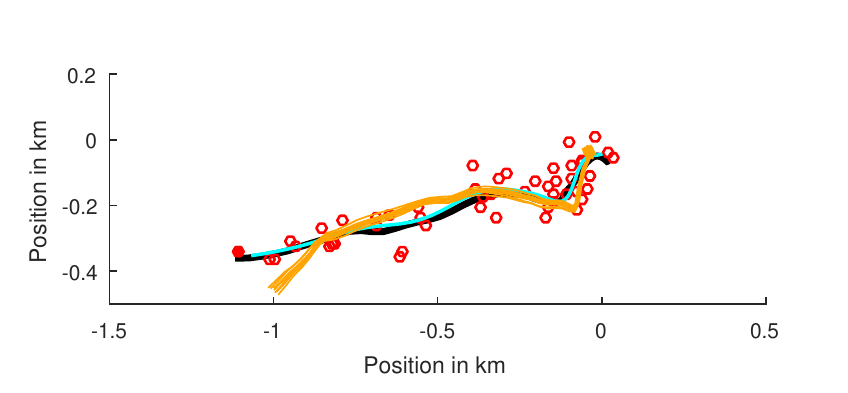}
  \caption{An ensemble with $N=10$ after updating with all measurements in the order $y_1, \dotsc y_L$. The smaller ensemble is more condensed and does not cover the RTS solution well.}
  \label{fig:batch_orderforward_N10_meas49_locdefaultInf}
\end{figure}

The next experiment explores the EnKF for $N=10$.
%
%
Fig.~\ref{fig:batch_orderforward_N10_meas49_locdefaultInf} shows the ensemble after processing all measurements. 
The ensemble is compactly gathered but does not cover the true trajectory well. The EnKF is overconfident.

A last experiment explores how well an EnKF with $N=20$ captures the uncertainty of the state estimate. Furthermore, we discuss effects of the order in which the measurements are processed. Specifically, we compare the ensemble covariance of the positions $\mathsf{x}_k$ to the exact $\cov(\mathsf{x}_k, \mathsf{x}_i)$, $i,k=0,\dotsc,L$, obtained by KF updates for the augmented state $\xi$. 

\begin{figure}
  \centering
  \includegraphics{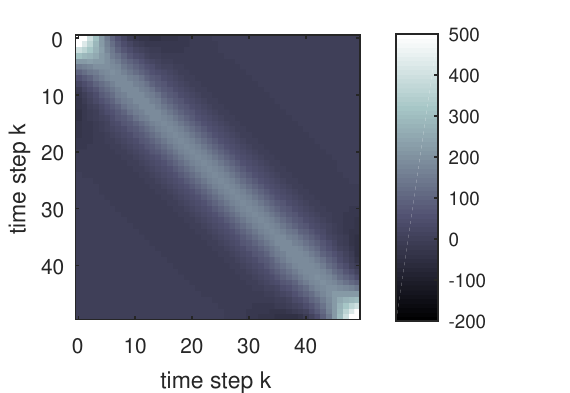}
  \caption{Exact position covariance matrix $\cov(\mathsf{x}_i, \mathsf{x}_j)$ after including all measurements.}
  \label{fig:batchcorr}
\end{figure}

The exact covariance after processing all measurements is illustrated in Fig.~\ref{fig:batchcorr}. 
Row $k$ in the matrix defines the covariance function between $\mathsf{x}_k$ and the remaining $\mathsf{x}$ positions.  
The banded structure indicates that subsequent positions are more related than, say, $\mathsf{x}_0$ and $\mathsf{x}_L$. 

\begin{figure}
  \centering
  \includegraphics{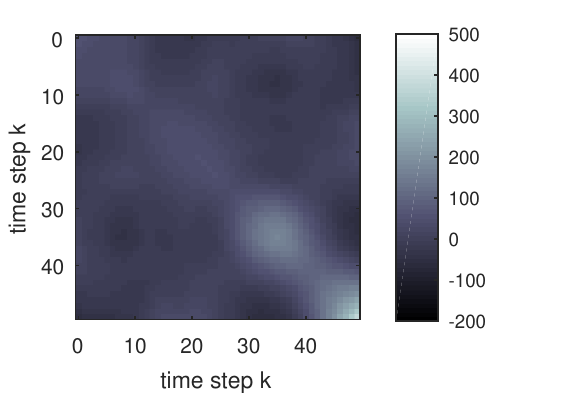}
  \caption{EnKF (20 members) position covariance matrix $\cov(\mathsf{x}_i, \mathsf{x}_j)$ after including all measurements in the order $y_1, \dotsc y_L$.}
  \label{fig:batchcorr_orderforward_N20_locdefaultInf}
\end{figure}

Fig.~\ref{fig:batchcorr_orderforward_N20_locdefaultInf} shows the corresponding EnKF covariance after processing the measurements from $y_1$ to $y_L$. 
The off-diagonal elements do not decay uniformly as in~Fig.~\ref{fig:batchcorr} and spurious positive and negative correlations appear. Furthermore, the correct temporal order of measurements entails an unwanted structure. Later $\mathsf{x}_k$ are rated more uncertain according to the lighter areas in the lower right corner of~Fig.~\ref{fig:batchcorr_orderforward_N20_locdefaultInf}.

\begin{figure}
  \centering
  \includegraphics{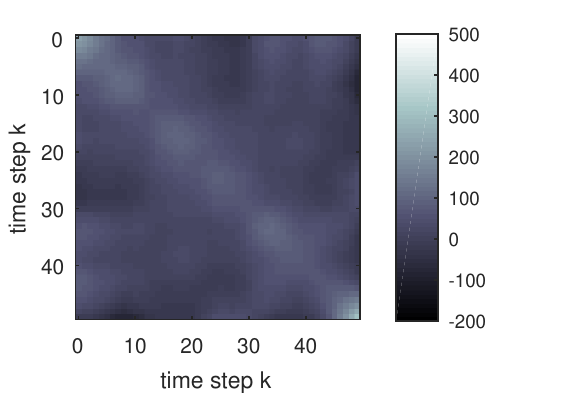}
  \caption{EnKF (20 members) position covariance matrix $\cov(\mathsf{x}_i, \mathsf{x}_j)$ after including all measurements in random order.}
  \label{fig:batchcorr_orderrandom_N20_locdefaultInf}
\end{figure}

A covariance after processing the measurements in random order is shown in Fig.~\ref{fig:batchcorr_orderrandom_N20_locdefaultInf}. 
The spurious correlations persist but the diagonal elements appear more homogeneous. 


From the above experiments we conclude that the EnKF can provide good estimates for ensembles with $N<n$. However, there is a minimum $N$ required to obtain consistent results without further measures such as localization or inflation. We have shown adverse effects such as ensembles with too little spread and spurious correlations. 

\subsection{The 40-dimensional Lorenz model}

\begin{figure}
  \centering
  \includegraphics{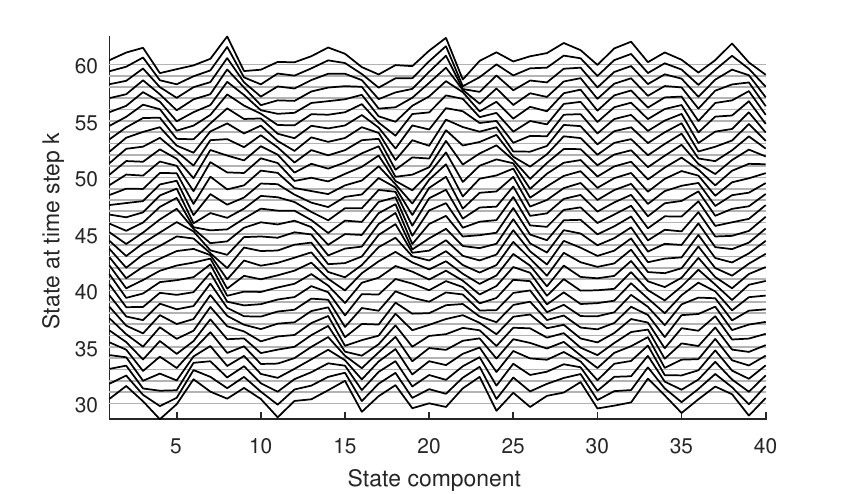}
  \caption{State evolution for the Lorenz model. Each horizontal line ``carries'' a 40-dimensional state vector.}
  \label{fig:lorenzStates}
\end{figure}

Our final example is a benchmark problem from the EnKF literature. We investigate the 40-dimensional Lorenz-96 model\footnote{Also known as the Lorenz-96, L95, L96, or L40 model.} from \cite{lorenz_predictability_2006} that is used in, e.g., \cite{anderson_ensemble_2001, whitaker_ensemble_2002, ott_local_2004, hunt_efficient_2007, sakov_implications_2008, van_leeuwen_nonlinear_2010, poterjoy_localized_2015}. 
The state $\mathsf{x}$ mimics an atmospheric quantity at equally spaced locations along a circle. Its evolution is specified by the nonlinear differential equation
\begin{align}
\dot{\mathsf{x}}(j) = \Bigl(\mathsf{x}(j+1)-\mathsf{x}(j-2)\Bigr) \mathsf{x}(j-1) - \mathsf{x}(j) + \mathsf{F}(j), \label{eq:lorenzODE}
\end{align}
where $j=1,\dotsc,40$ indexes the components of $\mathsf{x}$, with the convention that $\mathsf{x}(0)=\mathsf{x}(40)$ etc. Instead of the commonly used forcing term $\mathsf{F}(j)=8$, we assume time-dependent $\mathsf{F}\k(j)\sim\N(8,1)$ that are constant for time intervals $\mathsf{T}=0.05$ only and act as process noise. 
A Runge-Kutta method (RK4) is used to discretize~\eqref{eq:lorenzODE} to obtain the nonlinear state difference equation~\eqref{eq:stateDiffNonlin} with $x\k=\mathsf{x}\k$ and $v\k=\mathsf{F\k}$.
The step size $\mathsf{T}$ corresponds to about six hours if $\mathsf{x}$ were an atmospheric quantity on a latitude circle of the earth~\cite{lorenz_predictability_2006}.
Although the model~\eqref{eq:lorenzODE} is said to be chaotic, the effects are only mild for short integration times $\mathsf{T}$.
In our experiments all $n=40$ states are measured with additive Gaussian noise $e\k\sim\N(0,I)$. The initial state is Gaussian with $x_0\sim\N(0,P_0)$, where $P_0$ is drawn from a Wishart distribution with seed matrix $I_n$ and $n$ degrees of freedom. 
Fig.~\ref{fig:lorenzStates} illustrates how the state evolves over several time steps. 
There is a tendency for peaks to move ``westwards'' as $k$ increases.

We note that there are also alternative approaches for estimating $\mathsf{x}$, for example, by first linearizing and then discretizing~\eqref{eq:lorenzODE}. However, we adopt the RK4 discretization of the EnKF literature that yields a state transition that is easy to evaluate but difficult to linearize. Because of this, the EKF~\cite{anderson_optimal_1979} cannot be applied easily and we obtain a challenging benchmark problem.

We use sampling based EnKF to estimate long state sequences of $L=10^4$ time steps. 
Following~\cite{ott_local_2004,whitaker_ensemble_2002}, the performance is assessed by the error
\begin{equation}
\varepsilon\k = \sqrt{\frac{1}{n}(\xh\kk-x\k)\T (\xh\kk-x\k)},
\end{equation}
where $\xh\kk$ is the ensemble mean. 
We use the average $\varepsilon\k$ for $k=100,\dotsc,L$, denoted by $\bar\varepsilon$, as quantitative performance measure for different EnKF. Useful EnKF must yield $\bar\varepsilon<1$, which is the error when simply taking $\xh\kk=y\k$. 

\begin{figure}
  \centering
  \includegraphics{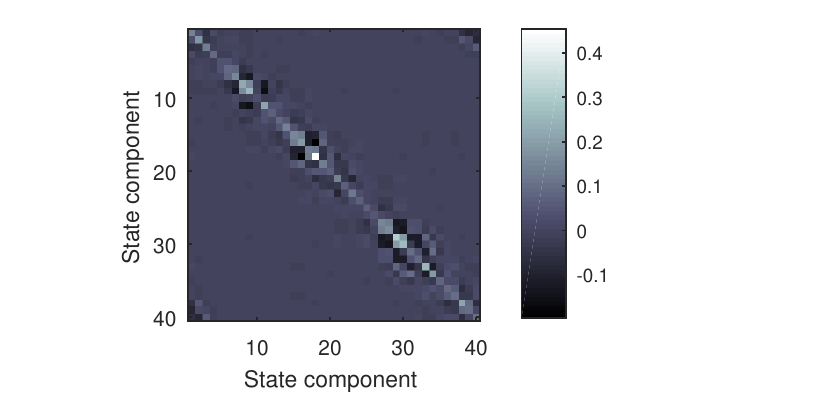}
  \caption{Prediction covariance $\bar P\kkm$ for $k=30$ obtained from an EnKF with $N=1000$. The banded structure justifies the use of localization.}
  \label{fig:lorenzCov}
\end{figure}

\begin{figure}
  \centering
  \includegraphics{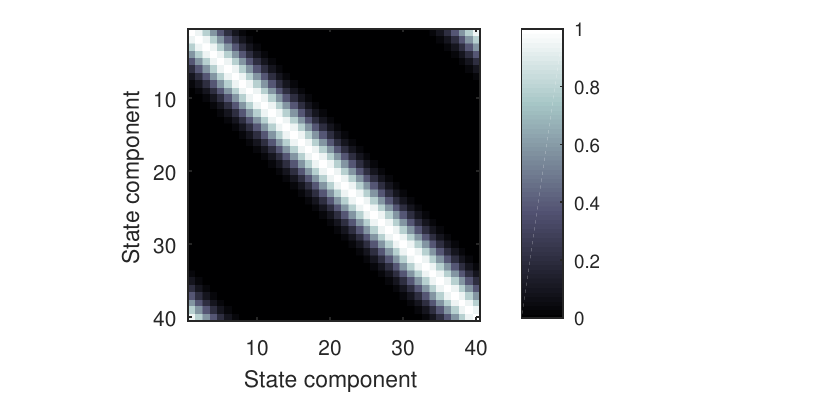}
  \caption{The employed tapering matrix $\rho$.}
  \label{fig:taperMat}
\end{figure}

First, we compute a reference solution using an EnKF with $N=1000$. Without any localization or inflation $\bar\varepsilon=0.29$ is achieved. Fig.~\ref{fig:lorenzCov} shows the sample covariance $\bar P\kkm$ of a prediction ensemble $X\kkm$, our best guess of the true covariance. 
The banded structure reveals that the problem is suitable for localization. 
Hence, we construct a matrix $\rho$ for covariance tapering from a compactly supported correlation function~\cite{gaspari_construction_1999} that is also used in~\cite{houtekamer_sequential_2001,hamill_distance-dependent_2001,whitaker_ensemble_2002,furrer_estimation_2007} and appears to be the standard choice. 
%
%
The chosen $\rho$ is a Toeplitz matrix because the components of $x\k$ are at equidistant locations, and shown in Fig.~\ref{fig:taperMat}. 

\begin{figure}
  \centering
  \includegraphics{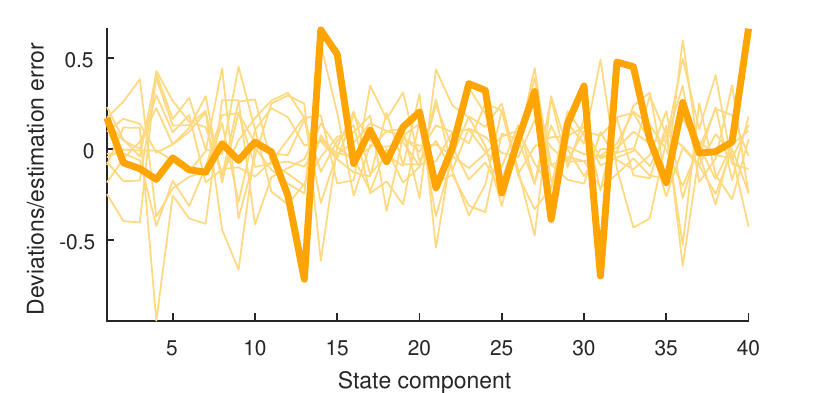}
  \caption{The estimation error $x\k-\bar x\kk$ for $k=10^4$ with the deviation ensemble $\widetilde X\kk$ in the background for an EnKF with $N=10$, covariance localization, and inflation factor $c=1.05$.}
  \label{fig:basic_N_10_inflation_5_kMax_10000_errVec}
\end{figure}

Next, EnKF with different ensemble sizes $N$, ensemble inflation factors $c$, with or without tapering, are compared. The obtained errors $\bar\varepsilon$ are summarized in Table~\ref{tab:results}.
\begin{table}[ht]
\centering
\caption{Averaged errors $\bar\varepsilon$ for different EnKF.}
\label{tab:results}
\begin{tabular}{l|l|l||l}
$N$ & $c$ & $\rho$ & $\bar\varepsilon$\\
\hline
$1000$ & $1$ & no & $0.29$\\
\hline
$40$ & $1$ & no & $0.44$\\
$40$ & $1.05$ & no & $0.33$\\
$40$ & $1$ & yes & $0.29$\\
$40$ & $1.02$ & yes & $0.28$\\
\hline
$20$ & $>1$ & no & $>1$\\
$20$ & $1.01$ & yes & $0.3$\\
\hline
$10$ & $1.05$ & yes & $0.34$\\
\end{tabular}
\end{table}
For $N=n=40$, we obtain a worse $\bar\varepsilon$ than for $N=1000$. While inflation without tapering does reduce the error slightly, the covariance tapering even yields a better result that the EnKF with $N=1000$. Further improvements are obtained by combining inflation and tapering. 
\begin{figure}
  \centering
  \includegraphics{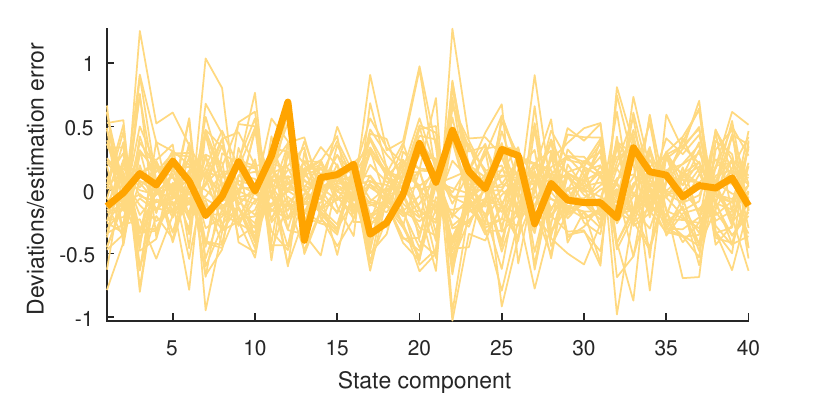}
  \caption{The estimation error $x\k-\bar x\kk$ for $k=10^4$ with the deviation ensemble $\widetilde X\kk$ in the background for an EnKF with $N=40$, covariance localization, and inflation factor $c=1.02$.}
  \label{fig:basic_N_40_inflation_2_kMax_10000_Q_1_gasp_errVec}
\end{figure}
Fig.~\ref{fig:basic_N_40_inflation_2_kMax_10000_Q_1_gasp_errVec} shows the estimation error $x\k-\xh\kk$ for $k=10^4$, $N=40$, $c=1.02$, and tapering with $\rho$. In the background, the ensemble deviations $\widetilde X\kk$ are illustrated. The estimation error is mostly contained in the intervals spanned by the ensemble, hence the EnKF is consistent. 
Tests on EnKF with $N=20$ reveal convergence problems, even with inflation the initial estimation error persists. With the help of tapering, however, a competitive error can be achieved. 
Even further reduction to $N=10$ is possible with tapering and inflation. The required inflation factor $c$ must be increased to counteract the lack of ensemble spread. Similar to Fig.~\ref{fig:basic_N_40_inflation_2_kMax_10000_Q_1_gasp_errVec}, Fig.~\ref{fig:basic_N_10_inflation_5_kMax_10000_errVec} illustrates the estimation error and deviation ensemble for $k=10^4$, $N=10$, $c=1.05$, and tapering with $\rho$. Although the obtained error is larger than for $N=40$, the ensemble deviations represent the estimation uncertainty well. 

A number of lessons have been learned from related experiments. As alternative to the $\rho$ in Fig.~\ref{fig:taperMat}, a simpler taper that contains only ones and zeros to enforce the banded structure was used. Although this $\rho$ was indefinite, a reduction in $\bar\varepsilon$ was achieved without any numerical issues. Hence, the specific structure of $\rho$ appears secondary. The smooth $\rho$ of Fig.~\ref{fig:taperMat} remains preferable in terms of $\bar\varepsilon$, though. 
Sequential processing of the measurements did not degrade the performance. 
Experiments without process noise give the lower errors $\bar\varepsilon$ from, e.g., \cite{ott_local_2004,whitaker_ensemble_2002}.

\section{Concluding Remarks}
\label{sec:conclusion}

With this paper, we have given a comprehensive and easy to understand introduction to the EnKF for signal processing researchers. The origin of the EnKF in the KF and its simple implementation have been demonstrated. The unique literature review provides quick access to the most relevant papers in the plethora of geoscientific EnKF publications. Furthermore, we have discussed the challenges related to small ensembles for high-dimensional states, $N<n$, and the available solutions such as localization or inflation. Finally, we have tested the EnKF on signal processing and EnKF benchmark problems. 

With its scalability and simple implementation, even for nonlinear and non-Gaussian problems, the EnKF stands out as viable candidate for many state estimation problems. Furthermore, localization ideas and advanced concepts for estimating covariance matrices and the EnKF gain from the limited information in the ensembles provide new research directions for the EnKF and high-dimensional filters in general, hopefully with an increased participation from the signal processing community.

\section*{Acknowledgments}

This work was supported by the project Scalable Kalman Filters granted by the Swedish Research Council (VR).



\bibliographystyle{IEEEtran}
\bibliography{IEEEabrv,enkfJournal}

\begin{thebibliography}{10}
\providecommand{\url}[1]{#1}
\csname url@samestyle\endcsname
\providecommand{\newblock}{\relax}
\providecommand{\bibinfo}[2]{#2}
\providecommand{\BIBentrySTDinterwordspacing}{\spaceskip=0pt\relax}
\providecommand{\BIBentryALTinterwordstretchfactor}{4}
\providecommand{\BIBentryALTinterwordspacing}{\spaceskip=\fontdimen2\font plus
\BIBentryALTinterwordstretchfactor\fontdimen3\font minus
  \fontdimen4\font\relax}
\providecommand{\BIBforeignlanguage}[2]{{%
\expandafter\ifx\csname l@#1\endcsname\relax
\typeout{** WARNING: IEEEtran.bst: No hyphenation pattern has been}%
\typeout{** loaded for the language `#1'. Using the pattern for}%
\typeout{** the default language instead.}%
\else
\language=\csname l@#1\endcsname
\fi
#2}}
\providecommand{\BIBdecl}{\relax}
\BIBdecl

\bibitem{kalnay_atmospheric_2002}
E.~Kalnay, \emph{\BIBforeignlanguage{English}{Atmospheric Modeling, Data
  Assimilation and Predictability}}.\hskip 1em plus 0.5em minus 0.4em\relax New
  York: Cambridge University Press, Dec. 2002.

\bibitem{kalman_new_1960}
R.~E. {{{{{Kalman}}}}}, ``A new approach to linear filtering and prediction
  problems,'' \emph{Journal of basic Engineering}, vol.~82, no.~1, pp. 35--45,
  Mar. 1960.

\bibitem{anderson_optimal_1979}
B.~D. Anderson and J.~B. Moore, \emph{Optimal Filtering}.\hskip 1em plus 0.5em
  minus 0.4em\relax Prentice Hall, Jun. 1979.

\bibitem{julier_new_1995}
S.~Julier, J.~Uhlmann, and H.~Durrant-Whyte, ``A new approach for filtering
  nonlinear systems,'' in \emph{Proceedings of the {American} {Control}
  {Conference} 1995}, vol.~3, 1995, pp. 1628--1632.

\bibitem{roth_nonlinear_2016}
M.~Roth, G.~Hendeby, and F.~Gustafsson, ``Nonlinear {{{{{Kalman}}}}} filters
  explained: A tutorial on moment computations and sigma point methods,''
  \emph{Journal of Advances in {I}nformation Fusion}, vol.~11, no.~1, pp.
  47--70, Jun. 2016.

\bibitem{gordon_novel_1993}
N.~J. Gordon, D.~J. Salmond, and A.~F. Smith,
  ``\BIBforeignlanguage{English}{Novel approach to
  nonlinear/non-{{{{{Gaussian}}}}} {{{{{Bayesian}}}}} state estimation},''
  \emph{\BIBforeignlanguage{English}{Radar and Signal Processing, {I}EE
  Proceedings F}}, vol. 140, no.~2, pp. 107--113, Apr. 1993.

\bibitem{gustafsson_particle_2010}
F.~Gustafsson, ``Particle filter theory and practice with positioning
  applications,'' \emph{IEEE Aerospace and Electronic Systems Magazine},
  vol.~25, no.~7, pp. 53--82, 2010.

\bibitem{evensen_sequential_1994}
G.~Evensen, ``\BIBforeignlanguage{en}{Sequential data assimilation with a
  nonlinear quasi-geostrophic model using {Monte} {Carlo} methods to forecast
  error statistics},'' \emph{\BIBforeignlanguage{en}{Journal of Geophysical
  Research: Oceans}}, vol.~99, no.~C5, pp. 10\,143--10\,162, 1994.

\bibitem{burgers_analysis_1998}
G.~Burgers, P.~Jan~van Leeuwen, and G.~Evensen, ``Analysis scheme in the
  ensemble {{{{{Kalman}}}}} filter,'' \emph{Monthly Weather Review}, vol. 126,
  no.~6, pp. 1719--1724, Jun. 1998.

\bibitem{durrant-whyte_simultaneous_2006}
H.~Durrant-Whyte and T.~Bailey, ``Simultaneous localization and mapping: {Part}
  {I},'' \emph{IEEE Robotics Automation Magazine}, vol.~13, no.~2, pp. 99--110,
  Jun. 2006.

\bibitem{baum_extended_2014}
M.~Baum and U.~D. Hanebeck, ``Extended object tracking with random hypersurface
  models,'' \emph{IEEE Transactions on Aerospace and Electronic Systems},
  vol.~50, no.~1, pp. 149--159, Jan. 2014.

\bibitem{wahlstrom_extended_2015}
N.~Wahlström and E.~Özkan, ``Extended target tracking using
  {{{{{Gaussian}}}}} processes,'' \emph{IEEE Transactions on Signal
  Processing}, vol.~63, no.~16, pp. 4165--4178, Aug. 2015.

\bibitem{houtekamer_data_1998}
P.~L. Houtekamer and H.~L. Mitchell, ``Data assimilation using an ensemble
  {{{{{Kalman}}}}} filter technique,'' \emph{Monthly Weather Review}, vol. 126,
  no.~3, pp. 796--811, Mar. 1998.

\bibitem{houtekamer_sequential_2001}
------, ``A sequential ensemble {{{{{Kalman}}}}} filter for atmospheric data
  assimilation,'' \emph{Monthly Weather Review}, vol. 129, no.~1, pp. 123--137,
  Jan. 2001.

\bibitem{jazwinski_stochastic_1970}
A.~H. Jazwinski, \emph{Stochastic Processes and Filtering Theory}.\hskip 1em
  plus 0.5em minus 0.4em\relax Academic Press, Mar. 1970.

\bibitem{evensen_ensemble_2003}
G.~Evensen, ``\BIBforeignlanguage{en}{The ensemble {{{{{Kalman}}}}} filter:
  theoretical formulation and practical implementation},''
  \emph{\BIBforeignlanguage{en}{Ocean Dynamics}}, vol.~53, no.~4, pp. 343--367,
  Nov. 2003.

\bibitem{evensen_data_2009}
------, \emph{\BIBforeignlanguage{English}{Data Assimilation: The Ensemble
  {{{{{Kalman}}}}} Filter}}, 2nd~ed.\hskip 1em plus 0.5em minus 0.4em\relax
  Dordrecht; New York: Springer, Aug. 2009.

\bibitem{hamill_ensemble-based_2006}
T.~M. Hamill, ``Ensemble-based atmospheric data assimilation,'' in
  \emph{Predictability of {Weather} and {Climate}}.\hskip 1em plus 0.5em minus
  0.4em\relax Cambridge University Press, 2006.

\bibitem{houtekamer_ensemble_2005}
P.~L. Houtekamer and H.~L. Mitchell, ``\BIBforeignlanguage{en}{Ensemble
  {{{{{Kalman}}}}} filtering},'' \emph{\BIBforeignlanguage{en}{Quarterly
  Journal of the Royal Meteorological Society}}, vol. 131, no. 613, pp.
  3269--3289, Oct. 2005.

\bibitem{whitaker_ensemble_2008}
J.~S. Whitaker, T.~M. Hamill, X.~Wei, Y.~Song, and Z.~Toth, ``Ensemble data
  assimilation with the {{{{{{{{{NCEP}}}}}}}}} global forecast system,''
  \emph{Monthly Weather Review}, vol. 136, no.~2, pp. 463--482, Feb. 2008.

\bibitem{compo_twentieth_2011}
G.~P. Compo, J.~S. Whitaker, P.~D. Sardeshmukh, N.~Matsui, R.~J. Allan, X.~Yin,
  B.~E. Gleason, R.~S. Vose, G.~Rutledge, P.~Bessemoulin, S.~Brönnimann,
  M.~Brunet, R.~I. Crouthamel, A.~N. Grant, P.~Y. Groisman, P.~D. Jones, M.~C.
  Kruk, A.~C. Kruger, G.~J. Marshall, M.~Maugeri, H.~Y. Mok, .~Nordli, T.~F.
  Ross, R.~M. Trigo, X.~L. Wang, S.~D. Woodruff, and S.~J. Worley,
  ``\BIBforeignlanguage{en}{The twentieth century reanalysis project},''
  \emph{\BIBforeignlanguage{en}{Quarterly Journal of the Royal Meteorological
  Society}}, vol. 137, no. 654, pp. 1--28, Jan. 2011.

\bibitem{lakshmivarahan_ensemble_2009}
S.~Lakshmivarahan and D.~Stensrud, ``Ensemble {{{{{Kalman}}}}} filter,''
  \emph{IEEE Control Systems}, vol.~29, no.~3, pp. 34--46, Jun. 2009.

\bibitem{anderson_ensemble_2009}
J.~Anderson, ``Ensemble {{{{{Kalman}}}}} filters for large geophysical
  applications,'' \emph{IEEE Control Systems}, vol.~29, no.~3, pp. 66--82, Jun.
  2009.

\bibitem{evensen_ensemble_2009}
G.~Evensen, ``The ensemble {{{{{Kalman}}}}} filter for combined state and
  parameter estimation,'' \emph{IEEE Control Systems}, vol.~29, no.~3, pp.
  83--104, Jun. 2009.

\bibitem{mandel_data_2009}
J.~Mandel, J.~Beezley, J.~Coen, and M.~Kim, ``Data assimilation for wildland
  fires,'' \emph{IEEE Control Systems}, vol.~29, no.~3, pp. 47--65, Jun. 2009.

\bibitem{furrer_estimation_2007}
R.~Furrer and T.~Bengtsson, ``Estimation of high-dimensional prior and
  posterior covariance matrices in {{{{{Kalman}}}}} filter variants,''
  \emph{Journal of Multivariate Analysis}, vol.~98, no.~2, pp. 227--255, Feb.
  2007.

\bibitem{butala_asymptotic_2008}
M.~Butala, J.~Yun, Y.~Chen, R.~Frazin, and F.~Kamalabadi, ``Asymptotic
  convergence of the ensemble {{{{{Kalman}}}}} filter,'' in \emph{15th {IEEE}
  {International} {Conference} on {Image} {Processing}}, Oct. 2008, pp.
  825--828.

\bibitem{mandel_convergence_2011}
J.~Mandel, L.~Cobb, and J.~D. Beezley, ``\BIBforeignlanguage{en}{On the
  convergence of the ensemble {{{{{Kalman}}}}} filter},''
  \emph{\BIBforeignlanguage{en}{Applications of Mathematics}}, vol.~56, no.~6,
  pp. 533--541, Dec. 2011.

\bibitem{frei_ensemble_2013}
M.~Frei, ``\BIBforeignlanguage{English}{Ensemble {{{{{Kalman}}}}} filtering and
  generalizations},'' Dissertation, ETH, Zürich, 2013, nr. 21266.

\bibitem{katzfuss_understanding_2016}
M.~Katzfuss, J.~R. Stroud, and C.~K. Wikle, ``Understanding the ensemble
  {{{{{Kalman}}}}} filter,'' \emph{The American Statistician}, pp. 350--357,
  Feb. 2016.

\bibitem{crisan_large_2011}
F.~Le~Gland, V.~Monbet, and V.~Tran, ``Large sample asymptotics for the
  ensemble {{{{{Kalman}}}}} filter,'' in \emph{The {Oxford} {Handbook} of
  {Nonlinear} {Filtering}}, D.~Crisan and B.~Rozovskii, Eds.\hskip 1em plus
  0.5em minus 0.4em\relax Oxford University Press, 2011, pp. 598--634.

\bibitem{butala_tomographic_2009}
M.~Butala, R.~Frazin, Y.~Chen, and F.~Kamalabadi, ``Tomographic {I}maging of
  dynamic objects with the ensemble {{{{{Kalman}}}}} filter,'' \emph{IEEE
  Transactions on {I}mage Processing}, vol.~18, no.~7, pp. 1573--1587, Jul.
  2009.

\bibitem{dunik_random-point-based_2015}
J.~Dunik, O.~Straka, M.~Simandl, and E.~Blasch, ``Random-point-based filters:
  analysis and comparison in target tracking,'' \emph{IEEE Transactions on
  Aerospace and Electronic Systems}, vol.~51, no.~2, pp. 1403--1421, Apr. 2015.

\bibitem{gillijns_what_2006}
S.~Gillijns, O.~Mendoza, J.~Chandrasekar, B.~De~Moor, D.~Bernstein, and
  A.~Ridley, ``What is the ensemble {{{{{Kalman}}}}} filter and how well does
  it work?'' in \emph{American {Control} {Conference}, 2006}, Jun. 2006, pp.
  4448--4453.

\bibitem{roth_ensemble_2015}
M.~Roth, C.~Fritsche, G.~Hendeby, and F.~Gustafsson, ``The ensemble
  {{{{{Kalman}}}}} filter and {I}ts relations to other nonlinear filters,'' in
  \emph{European {Signal} {Processing} {Conference} 2015 ({EUSIPCO} 2015)},
  Nice, France, Aug. 2015.

\bibitem{anderson_ensemble_2001}
J.~L. Anderson, ``An ensemble adjustment {{{{{Kalman}}}}} filter for data
  assimilation,'' \emph{Monthly Weather Review}, vol. 129, no.~12, pp.
  2884--2903, Dec. 2001.

\bibitem{bishop_adaptive_2001}
C.~H. Bishop, B.~J. Etherton, and S.~J. Majumdar, ``Adaptive sampling with the
  ensemble transform {{{{{Kalman}}}}} filter. {Part} {I}: Theoretical
  aspects,'' \emph{Monthly Weather Review}, vol. 129, no.~3, pp. 420--436, Mar.
  2001.

\bibitem{whitaker_ensemble_2002}
J.~S. Whitaker and T.~M. Hamill, ``Ensemble data assimilation without perturbed
  observations,'' \emph{Monthly Weather Review}, vol. 130, no.~7, pp.
  1913--1924, Jul. 2002.

\bibitem{tippett_ensemble_2003}
M.~K. Tippett, J.~L. Anderson, C.~H. Bishop, T.~M. Hamill, and J.~S. Whitaker,
  ``Ensemble square root filters,'' \emph{Monthly Weather Review}, vol. 131,
  no.~7, pp. 1485--1490, Jul. 2003.

\bibitem{anderson_monte_1999}
J.~L. Anderson and S.~L. Anderson, ``A {Monte} {Carlo} {I}mplementation of the
  nonlinear filtering problem to produce ensemble assimilations and
  forecasts,'' \emph{Monthly Weather Review}, vol. 127, no.~12, pp. 2741--2758,
  Dec. 1999.

\bibitem{van_leeuwen_comment_1999}
P.~J. van Leeuwen, ``Comment on “data assimilation using an ensemble
  {{{{{Kalman}}}}} filter technique”,'' \emph{Monthly Weather Review}, vol.
  127, no.~6, pp. 1374--1377, Jun. 1999.

\bibitem{ott_local_2004}
E.~Ott, B.~R. Hunt, I.~Szunyogh, A.~V. Zimin, E.~J. Kostelich, M.~Corazza,
  E.~Kalnay, D.~J. Patil, and J.~A. Yorke, ``\BIBforeignlanguage{en}{A local
  ensemble {{{{{Kalman}}}}} filter for atmospheric data assimilation},''
  \emph{\BIBforeignlanguage{en}{Tellus A}}, vol.~56, no.~5, Oct. 2004.

\bibitem{hamill_distance-dependent_2001}
T.~M. Hamill, J.~S. Whitaker, and C.~Snyder, ``Distance-dependent filtering of
  background error covariance estimates in an ensemble {{{{{Kalman}}}}}
  filter,'' \emph{Monthly Weather Review}, vol. 129, no.~11, pp. 2776--2790,
  Nov. 2001.

\bibitem{van_leeuwen_variance-minimizing_2003}
P.~J. van Leeuwen, ``A variance-minimizing filter for large-scale
  applications,'' \emph{Monthly Weather Review}, vol. 131, no.~9, pp.
  2071--2084, Sep. 2003.

\bibitem{snyder_obstacles_2008}
C.~Snyder, T.~Bengtsson, P.~Bickel, and J.~Anderson, ``Obstacles to
  high-dimensional particle filtering,'' \emph{Monthly Weather Review}, vol.
  136, no.~12, pp. 4629--4640, Dec. 2008.

\bibitem{van_leeuwen_particle_2009}
P.~J. van Leeuwen, ``Particle filtering in geophysical systems,'' \emph{Monthly
  Weather Review}, vol. 137, no.~12, pp. 4089--4114, Dec. 2009.

\bibitem{van_leeuwen_nonlinear_2010}
------, ``\BIBforeignlanguage{en}{Nonlinear data assimilation in geosciences:
  an extremely efficient particle filter},''
  \emph{\BIBforeignlanguage{en}{Quarterly Journal of the Royal Meteorological
  Society}}, vol. 136, no. 653, pp. 1991--1999, Oct. 2010.

\bibitem{frei_bridging_2013}
M.~Frei and H.~R. Künsch, ``\BIBforeignlanguage{en}{Bridging the ensemble
  {{{{{Kalman}}}}} and particle filters},''
  \emph{\BIBforeignlanguage{en}{Biometrika}}, vol. 100, no.~4, pp. 781--800,
  Dec. 2013.

\bibitem{poterjoy_localized_2015}
J.~Poterjoy, ``A localized particle filter for high-dimensional nonlinear
  systems,'' \emph{Monthly Weather Review}, vol. 144, no.~1, pp. 59--76, Oct.
  2015.

\bibitem{lorenz_predictability_2006}
E.~N. Lorenz, ``Predictability --- a problem partly solved,'' in
  \emph{Predictability of {Weather} and {Climate}}, T.~Palmer and R.~Hagedorn,
  Eds.\hskip 1em plus 0.5em minus 0.4em\relax Cambridge University Press, 2006,
  pp. 40--58.

\bibitem{pham_stochastic_2001}
D.~T. Pham, ``Stochastic methods for sequential data assimilation in strongly
  nonlinear systems,'' \emph{Monthly Weather Review}, vol. 129, no.~5, pp.
  1194--1207, May 2001.

\bibitem{luo_ensemble_2009}
X.~Luo and I.~Moroz, ``Ensemble {{{{{Kalman}}}}} filter with the unscented
  transform,'' \emph{Physica D: Nonlinear Phenomena}, vol. 238, no.~5, pp.
  549--562, Mar. 2009.

\bibitem{julier_unscented_2004}
S.~J. Julier and J.~K. Uhlmann, ``\BIBforeignlanguage{English}{Unscented
  filtering and nonlinear estimation},''
  \emph{\BIBforeignlanguage{English}{Proceedings of the {I}EEE}}, vol.~92,
  no.~3, pp. 401-- 422, Mar. 2004.

\bibitem{sakov_comment_2009}
P.~Sakov, ``Comment on “ensemble {{{{{Kalman}}}}} filter with the unscented
  transform”,'' \emph{Physica D: Nonlinear Phenomena}, vol. 238, no.~22, pp.
  2227--2228, Nov. 2009.

\bibitem{stordal_bridging_2011}
A.~S. Stordal, H.~A. Karlsen, G.~Nævdal, H.~J. Skaug, and B.~Vallès,
  ``\BIBforeignlanguage{en}{Bridging the ensemble {{{{{Kalman}}}}} filter and
  particle filters: the adaptive {{{{{Gaussian}}}}} mixture filter},''
  \emph{\BIBforeignlanguage{en}{Computational Geosciences}}, vol.~15, no.~2,
  pp. 293--305, Mar. 2011.

\bibitem{hoteit_particle_2011}
I.~Hoteit, X.~Luo, and D.-T. Pham, ``Particle {{{{{Kalman}}}}} filtering: A
  nonlinear {{{{{Bayesian}}}}} framework for ensemble {{{{{Kalman}}}}}
  filters,'' \emph{Monthly Weather Review}, vol. 140, no.~2, pp. 528--542, Aug.
  2011.

\bibitem{frei_mixture_2013}
M.~Frei and H.~R. Künsch, ``Mixture ensemble {{{{{Kalman}}}}} filters,''
  \emph{Computational Statistics \& Data Analysis}, vol.~58, pp. 127--138, Feb.
  2013.

\bibitem{trefethen_numerical_1997}
L.~N. Trefethen and {David Bau, {II}I},
  \emph{\BIBforeignlanguage{English}{Numerical Linear Algebra}}.\hskip 1em plus
  0.5em minus 0.4em\relax Philadelphia: SIAM, Jun. 1997.

\bibitem{norgaard_new_2000}
M.~Nørgaard, N.~K. Poulsen, and O.~Ravn, ``New developments in state
  estimation for nonlinear systems,'' \emph{Automatica}, vol.~36, no.~11, pp.
  1627--1638, Nov. 2000.

\bibitem{sakov_implications_2008}
P.~Sakov and P.~R. Oke, ``Implications of the form of the ensemble
  transformation in the ensemble square root filters,'' \emph{Monthly Weather
  Review}, vol. 136, no.~3, pp. 1042--1053, Mar. 2008.

\bibitem{livings_unbiased_2008}
D.~M. Livings, S.~L. Dance, and N.~K. Nichols, ``Unbiased ensemble square root
  filters,'' \emph{Physica D: Nonlinear Phenomena}, vol. 237, no.~8, pp.
  1021--1028, Jun. 2008.

\bibitem{lawson_implications_2004}
W.~G. Lawson and J.~A. Hansen, ``Implications of stochastic and deterministic
  filters as ensemble-based data assimilation methods in varying regimes of
  error growth,'' \emph{Monthly Weather Review}, vol. 132, no.~8, pp.
  1966--1981, Aug. 2004.

\bibitem{leeuwenburgh_impact_2005}
O.~Leeuwenburgh, G.~Evensen, and L.~Bertino, ``\BIBforeignlanguage{en}{The
  impact of ensemble filter definition on the assimilation of temperature
  profiles in the tropical pacific},'' \emph{\BIBforeignlanguage{en}{Quarterly
  Journal of the Royal Meteorological Society}}, vol. 131, no. 613, pp.
  3291--3300, Oct. 2005.

\bibitem{bar-shalom_estimation_2001}
Y.~Bar-Shalom, X.~R. Li, and T.~Kirubarajan, \emph{Estimation with Applications
  to Tracking and Navigation}.\hskip 1em plus 0.5em minus 0.4em\relax
  Wiley-Interscience, Jun. 2001.

\bibitem{sakov_relation_2010}
P.~Sakov and L.~Bertino, ``\BIBforeignlanguage{en}{Relation between two common
  localisation methods for the {EnKF}},''
  \emph{\BIBforeignlanguage{en}{Computational Geosciences}}, vol.~15, no.~2,
  pp. 225--237, Jul. 2010.

\bibitem{hunt_efficient_2007}
B.~R. Hunt, E.~J. Kostelich, and I.~Szunyogh, ``Efficient data assimilation for
  spatiotemporal chaos: A local ensemble transform {{{{{Kalman}}}}} filter,''
  \emph{Physica D: Nonlinear Phenomena}, vol. 230, no. 1–2, pp. 112--126,
  Jun. 2007.

\bibitem{gaspari_construction_1999}
G.~Gaspari and S.~E. Cohn, ``\BIBforeignlanguage{en}{Construction of
  correlation functions in two and three dimensions},''
  \emph{\BIBforeignlanguage{en}{Quarterly Journal of the Royal Meteorological
  Society}}, vol. 125, no. 554, pp. 723--757, Jan. 1999.

\bibitem{rasmussen_gaussian_2005}
C.~E. Rasmussen and C.~K.~I. Williams,
  \emph{\BIBforeignlanguage{English}{{{{{{Gaussian}}}}} Processes for Machine
  Learning}}.\hskip 1em plus 0.5em minus 0.4em\relax Cambridge, Mass: The MIT
  Press, Nov. 2005.

\bibitem{hastie_elements_2011}
T.~Hastie, R.~Tibshirani, and J.~Friedman,
  \emph{\BIBforeignlanguage{English}{The Elements of Statistical Learning: Data
  Mining, {I}nference, and Prediction}}, 2nd~ed.\hskip 1em plus 0.5em minus
  0.4em\relax New York, NY: Springer, Apr. 2011.

\bibitem{zhang_improving_2010}
Y.~Zhang and D.~S. Oliver, ``\BIBforeignlanguage{en}{Improving the ensemble
  estimate of the {{{{{Kalman}}}}} gain by bootstrap sampling},''
  \emph{\BIBforeignlanguage{en}{Mathematical Geosciences}}, vol.~42, no.~3, pp.
  327--345, Feb. 2010.

\bibitem{myrseth_resampling_2013}
I.~Myrseth, J.~Sætrom, and H.~Omre, ``Resampling the ensemble {{{{{Kalman}}}}}
  filter,'' \emph{Computers \& Geosciences}, vol.~55, pp. 44--53, Jun. 2013.

\bibitem{roth_computation_2017}
M.~Roth and F.~Gustafsson, ``Computation and visualization of posterior
  densities in scalar nonlinear and non-{{{{{Gaussian}}}}} {{{{{Bayesian}}}}}
  filtering and smoothing problems,'' in \emph{42nd {International}
  {Conference} on {Acoustics}, {Speech}, and {Signal} {Processing} ({ICASSP})},
  New Orleans, USA, Mar. 2017.

\end{thebibliography}

\end{document}